\RequirePackage{lineno} 
\documentclass[aps,prb,twocolumn,showpacs,showkeys,groupedaddress]{revtex4}  % for double-spaced preprint
\usepackage{graphicx}  % needed for figures
\usepackage{dcolumn}   % needed for some tables
\usepackage{bm}        % for math
\usepackage[english]{babel}
\usepackage{amssymb}   % for math
\usepackage[ansinew]{inputenc}
\usepackage{hyperref}
\usepackage{rotating}
\usepackage{multirow} 
\usepackage{float}
\usepackage{color}
\usepackage{pdfpages}

\begin{document}

%\setpagewiselinenumbers
%\modulolinenumbers[5]
%\linenumbers

%	\title{Measurements of quasi-particle tunneling in the second Landau level}
	\title{Experimental probe of topological orders and edge excitations in the second Landau level}
	
	\author{S.~Baer\footnote{Author to whom any correspondence should be addressed}, C.~R\"ossler, T.~Ihn, K.~Ensslin, C.~Reichl, W.~Wegscheider} 
	\affiliation{Solid State Physics Laboratory, ETH Z\"urich, 8093 Z\"urich, Switzerland}
	\email{sbaer@phys.ethz.ch}
	
	\date{\today}
	

	\begin{abstract}
%Numerical studies of the fractional quantum Hall states at $\nu$ = 7/3 and 8/3 have indicated that these states might not be well described by the Laughlin wave function. Subsequently, alternative wave functions with non-Abelian Quasiparticle (QP) excitations have been proposed, making these states, along with the 5/2 state, potentially interesting for topological quantum computation applications.
We measure weak quasiparticle tunneling across a constriction in the second Landau level. At $\nu$ = 7/3, 8/3 and 5/2, comparison of temperature and DC bias dependence to weak tunneling theory allows extracting parameters that describe the edges' quasiparticle excitations. 
At $\nu$ = 8/3, our results are well described by a particle-hole conjugate Laughlin state, but not compatible with proposed non-Abelian quasiparticle excitations. For $\nu$ = 5/2, our measurements are in good agreement with previous experiments and favor the Abelian {(3,3,1) or (1,1,3)-states}. At these filling factors, we further investigate the influence of the backscattering strength on the extracted scaling parameters. For $\nu$ = 7/3, the backscattering strength strongly affects the scaling parameters, whereas quasiparticle tunneling at $\nu$ = 8/3 and 5/2 appears more robust. 
Our results provide important additional insight about the physics in the second Landau level and contribute to the understanding of the physics underlying the fractional quantum Hall states at $\nu$ = 7/3, 8/3 and 5/2. 

	\end{abstract}

\maketitle

\section{Introduction}
\begin{figure}
	\begin{center}
	\includegraphics[width=8.6cm]{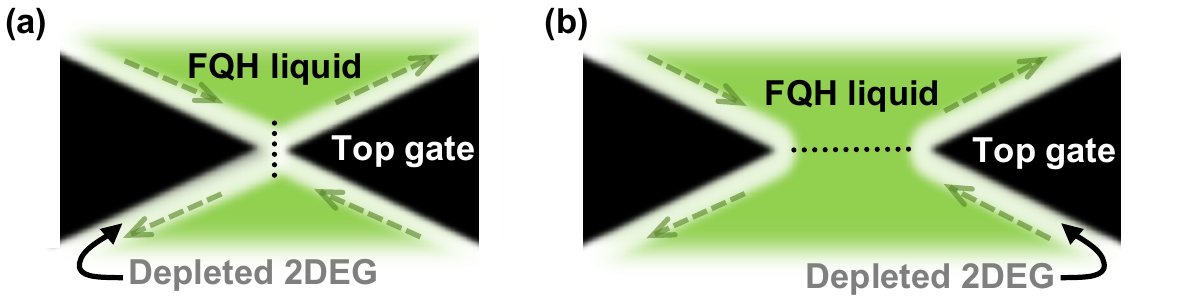}
	\end{center}
	\caption{(Color online) Conceptual difference between weak and strong backscattering \cite{chang_chiral_2003,wen_quantum_2007,fradkin_field_2013}, in the simplest case without edge reconstruction. \textbf{a}: For a quantum point contact (QPC) close to pinch-off, we have strong backscattering and weak electron tunneling (dotted line) . \textbf{b}: For an open QPC, weak backscattering and weak quasiparticle tunneling (dotted line) govern the transmission.}
	\label{QPCScheme}
	\end{figure}

Numerical studies of the fractional quantum Hall (FQH) states at $\nu$ = 7/3 and 8/3 have indicated that these states might not be well described by the Laughlin wave function \cite{wojs_electron_2001,balram_role_2013,macdonald_collective_1986,dAmbrumenil_fractional_1988}. Thus, the underlying physics which creates the energy gap might be different for $\nu$=1/3 and $\nu$=7/3 and 8/3. 
Subsequently, alternative wave functions with non-Abelian quasiparticle (QP) excitations have been proposed for $\nu$= 7/3 and 8/3 \cite{bonderson_fractional_2008,read_beyond_1999}, making these states, along with the 5/2 state \cite{willett_observation_1987,willett_quantum_2013,pan_exact_1999,eisenstein_insulating_2002}, potentially interesting for topologically protected quantum operations \cite{freedman_p/np_1998,kitaev_fault-tolerant_2003,nayak_non-abelian_2008}.\\
{Most current} experimental findings for both the $\nu$ = 7/3 and 8/3 states are compatible with non-Abelian candidate states and a (particle-hole conjugate) Laughlin state. For instance, local electrometer \cite{venkatachalam_local_2011} and shot noise measurements \cite{dolev_characterizing_2011,gross_upstream_2012} suggest a QP charge $e^*/e$ = 1/3. The latter experiments furthermore show that a neutral mode is present for $\nu$ = 8/3 but absent for $\nu$ = 7/3 . 
{From activation measurements, the $\nu$=7/3 and 8/3 states were found to be consistent with Jain's non-interacting composite fermion model \cite{kumar_nonconventional_2010}, hence supporting a (particle-hole conjugate) Laughlin state.
Nevertheless,} further experiments are necessary, which allow {a more direct discrimination of} the proposed wave functions. \\
Tunneling experiments employing {quantum point contacts (QPCs)} \cite{milliken_indications_1996} or structures made by cleaved-edge overgrowth \cite{chang_observation_1996} have been used to study the characteristic power-law scaling of the chiral Luttinger liquid tunneling conductance: 	
a $\nu$ = 1/3 edge was weakly tunneling-coupled to another FQH edge or to a bulk metal across vacuum.
Thus measured conductances arose from the tunneling of electrons (Fig. \ref{QPCScheme}.a, dotted line), which is strongly suppressed at low temperatures.
In the case where counterpropagating edge states are weakly coupled across a FQH liquid  (in the simplest case without edge reconstruction, Fig. \ref{QPCScheme}.b, dotted line), QPs tunnel between the edges \cite{chang_chiral_2003,wen_quantum_2007,fradkin_field_2013}. In contrast to the previous case, this process is strongly enhanced at low $T$. 
Weak QP tunneling has been used as a probe for edge properties of the $\nu$ = 5/2 state \cite{radu_quasi-particle_2008,lin_measurements_2012}. This situation recently also has been studied theoretically \cite{fendley_edge_2007,feiguin_nonequilibrium_2008,das_effect_2009,yang_influence_2013}.
The DC bias and temperature dependence of the tunneling conductance across a QPC was employed to extract the QP charge $e^*/e$ and the Coulomb interaction parameter $g$, which describes the strength of electron-electron interaction in a FQH edge and reflects the topological order in the bulk \cite{wen_theory_1992}. 
These parameters characterize the edge excitations of proposed wave functions for $\nu$ = 5/2, 7/3 and 8/3 and hence allow probing the nature of these states experimentally.\\
In this article, we use this technique for the investigation of the most prominent filling factors of the lower spin branch of the second Landau level (LL): $\nu$ = 7/3, 8/3 and 5/2. 
To the best of our knowledge, our results constitute the first detailed experimental investigation of scaling parameters $g$ and $e^*/e$ for the 7/3 and 8/3 state\footnote{In Ref. \onlinecite{yiming__zhang_waves_2009}, two groups of zero bias peaks were observed for $7/3 < \nu < 8/3$. These were attributed to $\nu=5/2$ and $\nu=8/3$. At $\nu=8/3$ the data was not conclusive, whereas for $\nu=5/2$ similar conclusions as in Ref. \onlinecite{radu_quasi-particle_2008} were reached.}. We provide a comparison to theoretical proposals.
At $\nu$ = 5/2, our extracted scaling parameters are very similar to those reported earlier \cite{radu_quasi-particle_2008,lin_measurements_2012}, though measured in a quantum well with a different growth technique and an approximately 12 $\%$ lower electron sheet density.
Finally, we study the effect of the backscattering strength of the QPC on the QP tunneling and the extracted parameters, and investigate the breakdown of weak QP tunneling.

\section{Experimental details}
The measured QPCs {are approximately 1.1 $\mu$m wide and} are defined by electron-beam lithography and subsequent Ti/Au evaporation on photo-lithographically patterned high-mobility wafers. These high mobility structures ($n_\mathrm{s}\approx 2.3 \times 10^{11}~\mathrm{cm}^{-2},~\mu \approx 2.3 \times 10^{7}~\mathrm{cm}^2/\mathrm{Vs}$) are optimized for the observation of the $\nu=5/2$ state without prior LED illumination \cite{reichl_increasing_2014}. {The 27 nm wide quantum well lies approximately 200 nm below the surface. A DX doping scheme has been used.}
Experiments have been conducted in a cryogen-free dilution refrigerator, with an electronic base temperature $T_\mathrm{el}$ $\approx$ 12 - 13 mK, achieved by low-pass filtering and thermal anchoring at every temperature stage. The bath temperature ($T_\mathrm{bath}\approx10$ mK) is measured with a SQUID-based noise thermometer, which gives reliable results down to temperatures below 10 mK \cite{engert_noise_2012,engert_practical_2009}. Top-gated structures have been cooled down from room temperature to 4 K with a positive pre-bias. Subsequently, top-gates have been negatively biased at 4 K to allow for density relaxation in the screening layers and the QPC channel \cite{rossler_gating_2010,radu_quasi-particle_2008,baer_interplay_2014}. The electron gas underneath the top-gates is depleted at a gate voltage of -1.4 V.
	\begin{figure}
	\begin{center}
	\includegraphics[width=8.6cm]{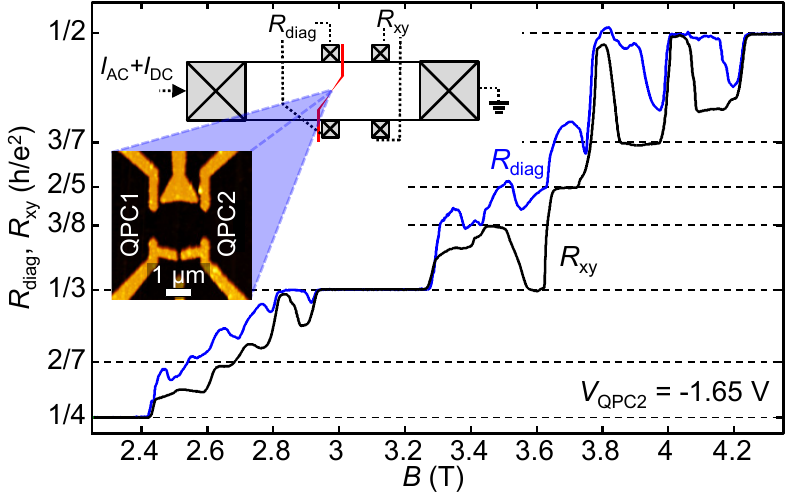}
	\end{center}
	\caption{(Color online) $R_\mathrm{diag}$ [blue (gray)] and $R_\mathrm{xy}$ (black) measured in a Hall-bar geometry (upper inset) as a function of the magnetic field. Here, -1.65 V have been applied to QPC2 (left inset). In between integer filling factors, $R_\mathrm{diag}\geq R_\mathrm{xy}$, indicating a reduced transmission of the QPC.}
	\label{BulkDiag}
	\end{figure}

\section{Measurement results}
Fig. \ref{BulkDiag} shows the bulk Hall resistance $R_\mathrm{xy}$ measured far away from the top-gate defined QPCs and the resistance measured diagonally across one of the QPCs, $R_\mathrm{diag}$, for filling factors in the bulk $2~\le~\nu_\mathrm{bulk}~\le~4$ at base temperature. 
Here, the QPC2 gates (see inset of Fig. \ref{BulkDiag}) are biased to -1.65 V (at the onset of weak quasiparticle tunneling), while all other gates are grounded. A constant AC current $I_\mathrm{AC}$ = 1.0 nA is applied at $f$ = 13.333 Hz, while $I_\mathrm{DC}$ = 0.
$R_\mathrm{diag}$ and $R_\mathrm{xy}$ are measured in a standard 4-terminal configuration (see inset of Fig. \ref{BulkDiag}) using lock-in measurement techniques.
In addition to the integer quantum Hall (IQH) states, FQH states at $\nu$ = 8/3, 5/2, 7/3 and strong reentrant integer quantum Hall (RIQH) states are observed in the bulk. 
Whenever an IQH plateau is observed in $R_\mathrm{xy}$, $R_\mathrm{diag}$ is quantized at exactly the same resistance value, indicating very similar bulk and QPC electronic densities. 
In-between the IQH plateaus, $R_\mathrm{diag} \ge R_\mathrm{xy}$, indicating reduced transmission through the QPC.
In this situation, weak backscattering of edge states through the QPC occurs via weak QP tunneling between counter-propagating edge states (Fig. \ref{QPCScheme}.b).
We measure the tunneling conductance across the QPC, $g_\mathrm{tun}\approx (R_\mathrm{diag}-R_\mathrm{xy})/R_\mathrm{xy}^2$ [\onlinecite{radu_quasi-particle_2008}] for different bulk filling factors $\nu_\mathrm{bulk}$. 
The power-law temperature dependence of the zero-bias tunneling conductance \cite{wen_edge_1991,wen_theory_1992} $g_\mathrm{tun}\vert_{I_\mathrm{{SD}=0}} \propto T^{2g-2}$ then allows extracting the Coulomb interaction parameter $g$, which can be compared to theoretical predictions. 
With an additional DC bias between the counter-propagating edges, the tunneling conductance takes the form \cite{wen_edge_1991,de_c._chamon_resonant_1993,de_c._chamon_two_1997,bas_jorn_overbosch_edge_2008}
\begin{equation}
g_\mathrm{tun} = A\times T^{(2g-2)}\times F\left(g,\frac{e^*/e~I_\mathrm{DC}R_\mathrm{xy}}{k_BT}\right)~+~g_\infty,
\label{TunnelFormula}
\end{equation}
{Here, a heuristic background conductance $g_\infty$ has been introduced. $F$ is a function} of $g$ and $\left(e^*/e~I_\mathrm{DC}R_\mathrm{xy}\right)/(k_BT)$ [\onlinecite{radu_quasi-particle_2008}]:
{
\begin{eqnarray*}
F\left(g,x\right)=&\mathrm{B}\left(g+i\frac{x}{2\pi},g-i\frac{x}{2\pi}\right) \times \\
 &\left\lbrace \pi \cosh\left(\frac{x}{2}\right) -2 \sinh \left(\frac{x}{2}\right)\mathrm{Im}\left[ \Psi \left(g+i\frac{x}{2\pi}\right)\right]\right\rbrace
\end{eqnarray*}}
{Here, $\mathrm{B}(x,y)$ is the Euler beta function and $\Psi(x)$ is the digamma function. A derivation of this formula can be found for example in References \onlinecite{de_c._chamon_two_1997,bas_jorn_overbosch_edge_2008}}

{This formula is the result of a perturbative calculation which assumes a point-like interaction of the counter-propagating edge states in the QPC \cite{wen_edge_1991,de_c._chamon_resonant_1993,de_c._chamon_two_1997,bas_jorn_overbosch_edge_2008}. It relies on the scaling dimensions of the most relevant quasiparticle creation and annihilation operators of the individual edges. The exact form of these operators depends on the FQH edge modes and their interactions. Edge theories and corresponding quasiparticle operators have been developed for all the relevant candidate wavefunctions in the second LL (an overview can be found for example in reference \onlinecite{yang_influence_2013}). As long as the interaction between the counter-propagating edge modes is weak and can be treated in a perturbative approach, we expect Eq. \ref{TunnelFormula} to be valid.}

Measuring the full DC bias and temperature dependence of the tunneling conductance gives access to $g$ and $e^*/e$ via comparsion to Eq. \ref{TunnelFormula}.

In the following, QP tunneling is studied in different configurations. First, the $B$-field is fixed to the center of the bulk filling factors and the QPC transmission is kept constant (similar to Refs. \onlinecite{radu_quasi-particle_2008,lin_measurements_2012}). In this configuration, $\nu=5/2$ (section \ref{52Const}) and $\nu=8/3$ (section \ref{83Const}) are investigated. Backscattering for $\nu=7/3$ is much weaker than for $\nu=5/2$ and $\nu=8/3$. For the QPC voltages chosen, a reliable parameter extraction was not possible for $\nu = 7/3$ (data not shown).
In section \ref{Bvar}, the influence of the magnetic field strength on the tunneling parameters is investigated. Finally, the influence of the QPC transmission is investigated (section \ref{QPCvar}). In the latter two sections, also backscattering at $\nu=7/3$ is observed in narrow parameter windows.

\subsection{Tunneling conductance at \texorpdfstring{$\nu$}{\textbackslash nu} = 5/2}
\label{52Const}
\begin{figure}[H]
	\begin{center}
	\includegraphics[width=8.6cm]{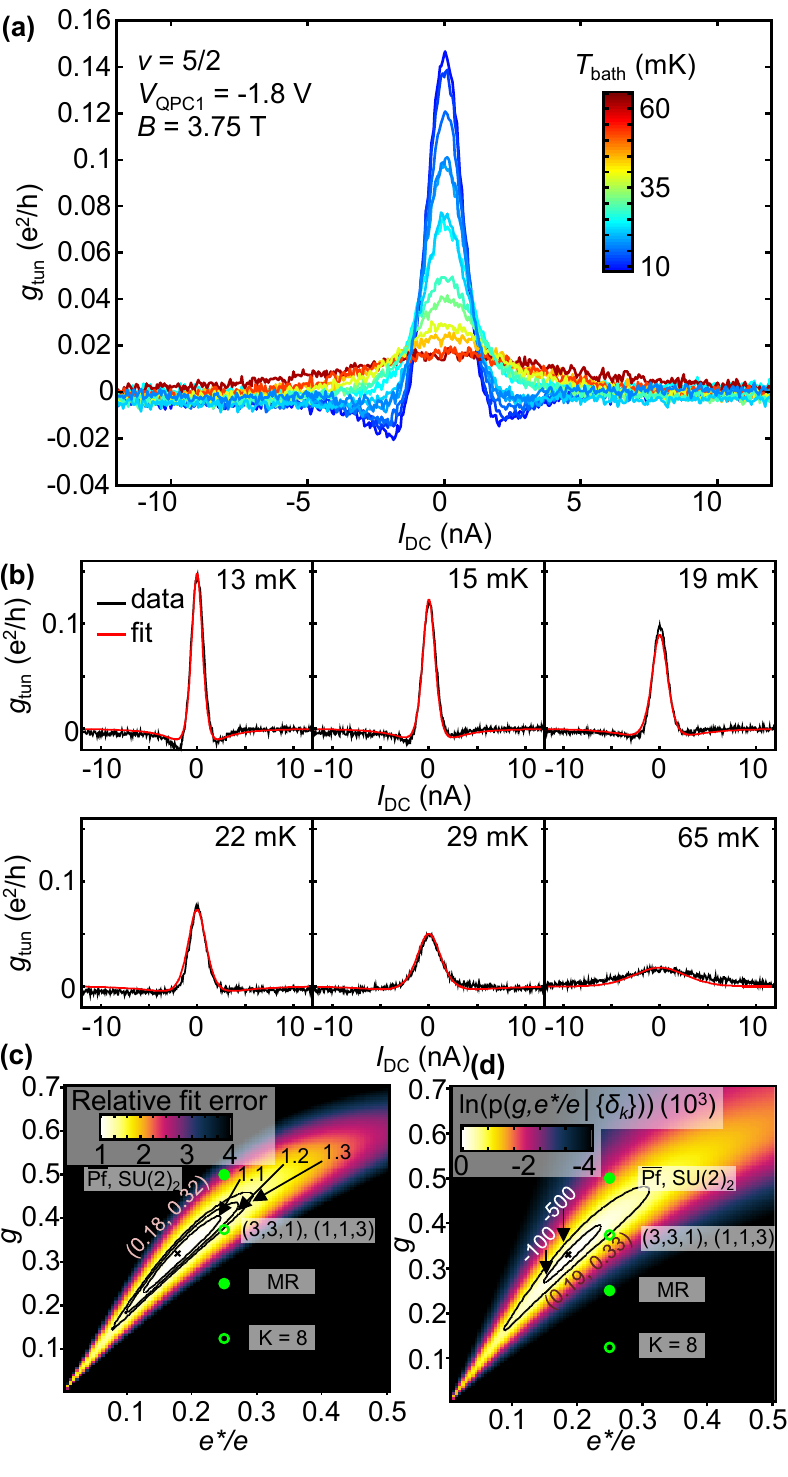}
	\end{center}
	\caption{(Color online) \textbf{a:} Zero-bias backscattering peak at $\nu$=5/2 and fixed $V_\mathrm{QPC1}$ = -1.8 V. The peak height is strongly temperature-dependent. \textbf{b:} Measured (black) and fitted (red) tunneling conductance for different electronic temperatures (fit parameters: $e^*/e$ = 0.18, $g$ = 0.32). \textbf{c:} Normalized fit error as a function of fit parameters $e^*/e$ and $g$. Parameters of Abelian [(green) circles] and non-Abelian [(green) dots] candidate wave functions are indicated. \textbf{d:} PDF of the measured residuals $\{\delta_k \}$ as a function of $e^*/e$ and $g$. The maximum probability is found for $e^*/e$ = 0.19, $g$ = 0.33 with $\sigma_\mathrm{g}$ = 0.0026, $\sigma_\mathrm{e^*/e}$ = 0.0019 and  $\sigma_\mathrm{g,e^*/e}$ = 0.0022}
	\label{TDep}
	\end{figure}

Fig. \ref{TDep}.a shows the temperature dependence of the measured $g_\mathrm{tun}$ of QPC1 when $V_\mathrm{QPC1}$ is fixed to -1.8 V. At this gate voltage, backscattering is sufficiently strong to be observed up to temperatures of $\approx$ 65 mK. {A narrow peak of the tunneling conductance is observed at zero DC current. Adjacent to the QP tunneling peak, undershoots of the tunneling conductance are observed. Such undershoots of the tunneling conductance are only expected for $g<$ 0.5 \cite{bas_jorn_overbosch_edge_2008,lin_measurements_2012}.}
The $B$-field is set to the center of the bulk $\nu$ = 5/2 plateau for this measurement, and an AC current $I_\mathrm{AC}$ = 0.4 nA is applied. Decreasing the AC current below this value does not narrow the $g_\mathrm{tun}$ peak, but only reduces the signal to noise ratio. 
A small constant background of approx. $0.1 \times e^2/h$ is removed from the data, by subtracting the saturation $g_\mathrm{tun}$ at $I_\mathrm{DC}\geq 10$ nA. When the temperature is increased to approx. 65 mK, the zero-bias peak vanishes almost completely. 
A fit of the weak tunneling expression (Eq. \ref{TunnelFormula}) to the measured $g_{\mathrm{tun}}$ is shown in Fig. \ref{TDep}.b (six out of 13 measured temperatures are shown). 
The parameters $g_\infty$, $A$, $g$ and $e^*/e$ are identical for all $T$ and are fitted to the data. With $e^*/e$ = 0.18 and $g$ = 0.32, excellent agreement of experiment and QP tunneling theory is obtained. These parameters are close to those reported in Refs. \onlinecite{radu_quasi-particle_2008,lin_measurements_2012}. There, best fit parameters $e^*/e$ = 0.17, $g$ = 0.35 \cite{radu_quasi-particle_2008} and $e^*/e$ = 0.25 / 0.22, $g$ = 0.42 / 0.34 (Ref.  \onlinecite{lin_measurements_2012}, for two different geometries) were found.
Suitable parameter ranges can be deduced from the fit residuals $\delta_k$ of the $k^\mathrm{th}$ measurement point.
We plot the relative fit error, i.e. $\Sigma_\delta=\min\limits_{A,g_\infty} \left(\sum\limits_{k} {\delta_k^2}\right)$ (normalized by its minimum, $\Sigma_{\delta,\mathrm{min}}$), as a function of $e^*/e$ and $g$ (Fig. \ref{TDep}.c), similar as it has been done in Refs. \onlinecite{radu_quasi-particle_2008,lin_measurements_2012}.
With this plot, the agreement with parameters for proposed wave functions can be assessed qualitatively. 
Parameters for different wave functions are cited in Table \ref{TabelleParameters} and are indicated in Figs. \ref{TDep}.c,d as (green) circles {(Abelian modes) or (green) dots (non-Abelian modes)}. All QP excitations are expected to possess a minimum $e^*/e$ = 0.25. The Abelian $K=8$ state \cite{wen_ground-state_1990,blok_effective_1990,wen_classification_1992} with $g$ = 0.125 does clearly not agree well with our experimental observations. {Very recently, it was shown that the (1,1,3)-state is also a viable candidate for $\nu$=5/2 \cite{yang_experimental_2014}. It is Abelian and is expected to possess $g\approx$0.375 in a gate-defined geometry. Closest agreement of our data seems to be found with this (1,1,3)-state and the Abelian (3,3,1)-state \cite{wen_classification_1992,yang_hierarchical_1992,blok_effective_1990} for which $g$ = 0.375 is expected.} The parameters of this state reproduce the experimental $g_\mathrm{tun}$ qualitatively well (see Supplementary Information).    The non-Abelian Moore-Read Pfaffian \cite{moore_nonabelions_1991} ($g$ = 0.25), Anti-Pfaffian \cite{levin_particle-hole_2007,lee_particle-hole_2007} ($\overline{\mathrm{Pf}}$, $g$ = 0.5), $\mathrm{SU(2)_2}$ state \cite{wen_non-abelian_1991,blok_many-body_1992}($g$ = 0.5) {and Majorana gapped edge-reconstructed Pfaffian state \cite{overbosch_phase_2008} ($g$=0.5)} seem less likely and also do not fit as well qualitatively (see Supplementary Information), though they cannot be excluded completely. {For the Majorana-gapped anti-Pfaffian \cite{overbosch_phase_2008} and the particle hole conjugate states, $\overline{\mathrm{(3,3,1)}}$ and $\overline{\mathrm{SU(2)_2}}$ \cite{yang_influence_2013}, $g>$ 0.5 is expected and they hence are not indicated in Fig. \ref{TDep}.c,d.}
For the best fit, $\chi^2=\Sigma_{\delta,\mathrm{min}}/(N \sigma_\mathrm{meas}^2)=2.14$ is found, where $N$ is the number of measurement points and $\sigma_\mathrm{meas}^2$ is the measurement noise (variance). This indicates a slight systematic disagreement between measurements and model function.

A more quantitative assessment can be gained from the probability distribution for $g$ and $e^*/e$, $p\left(g,e^*/e\vert\{\delta_k\}\right)$, which is calculated from the Gaussian probability density function (PDF) of our fit residuals, leading to the posterior probability $p\left(\sigma,A,g_\infty,g,e^*/e\vert \{\delta_k\}\right)$ by marginalization of the variables $\sigma,~A$ and $g_\infty$ \cite{sivia_data_2006}. 
The maximum probability is found for $e^*/e$ = 0.19 and $g$ = 0.33 with very narrow standard deviations $\sigma_g$ = 0.0026, $\sigma_{e^*/e}$ = 0.0019. 
The positive covariance $\sigma_{g,e^*/e}$ = 0.0022 indicates that we are more sensitive to the difference $g-e^*/e$ than to the individual parameters $g$ and $e^*/e$. 
Overall, we can conclude that our measurements are well described by the weak tunneling expression of equation \ref{TunnelFormula}, with only a small systematic deviation. However, when comparing the ``best fit" parameters to the proposed parameter pairs, the small size of the standard deviations suggest that there is clearly a systematic deviation.  Here, none of the proposed parameter pairs lie within our statistical error.

\subsection{Tunneling conductance at \texorpdfstring{$\nu$}{\textbackslash nu} = 8/3}
\label{83Const}
\begin{figure}[H]
	\begin{center}
	\includegraphics[width=8.6cm]{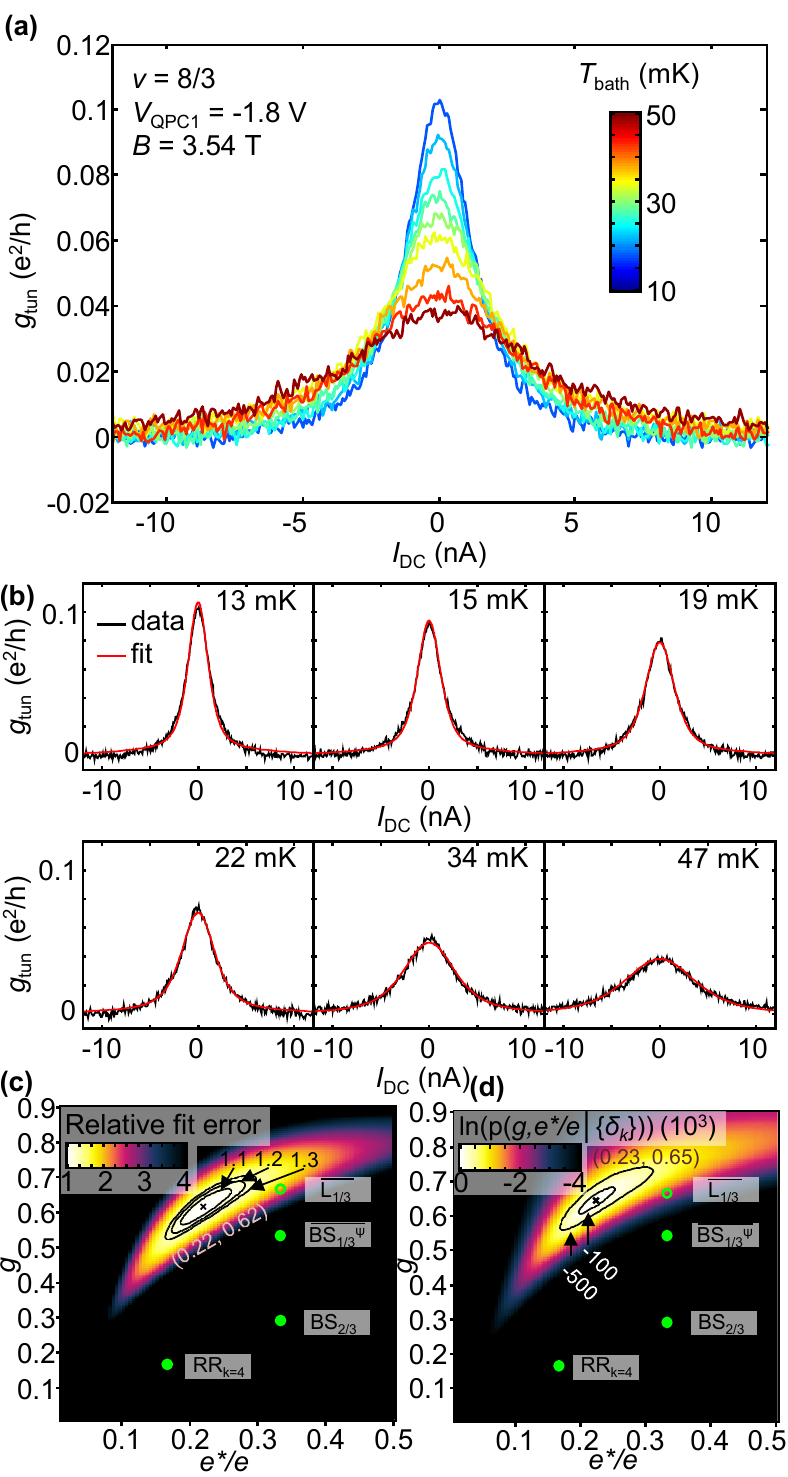}
	\end{center}
	\caption{(Color online) \textbf{a:} Zero-bias backscattering peak at $\nu$=8/3 and fixed $V_\mathrm{QPC1}$ = -1.8 V. \textbf{b:} Measured (black) and fitted (red) tunneling conductance for different electronic temperatures (fit parameters: $e^*/e$ = 0.22, $g$ = 0.62). \textbf{c:} Relative fit error as a function of fit parameters $e^*/e$ and $g$. Parameters of Abelian [(green) circles] and non-Abelian [(green) dots] candidate wave functions are indicated. \textbf{d:} PDF of the measured residuals $\{\delta_k\}$ as a function of $e^*/e$ and $g$. The maximum probability is found for $e^*/e$ = 0.23, $g$ = 0.65 with $\sigma_\mathrm{g}$ = 0.0029, $\sigma_\mathrm{e^*/e}$ = 0.0028 and  $\sigma_\mathrm{g,e^*/e}$ = 0.0028}
	\label{TDep83}
	\end{figure}

A similar analysis can be conducted at a bulk filling factor $\nu_\mathrm{bulk}$=8/3. For this the $B$-field is set to the center of the bulk 8/3 plateau. Here, the tunneling peak has a qualitatively different shape (Fig. \ref{TDep83}.a) with a larger full width at half maximum (FWHM) in $I_\mathrm{DC}$ direction and absent $g_\mathrm{tun}$ undershoots. {The absence of these undershoots is a sign for $g>$ 0.5 \cite{lin_measurements_2012,bas_jorn_overbosch_edge_2008}.}
Also in this case, the weak tunneling expression (Eq. \ref{TunnelFormula}) fits the data well over a large temperature range (six out of nine measured temperatures are shown in Fig. \ref{TDep83}.b). From the fit we obtain: $e^*/e$ = 0.22, $g$ = 0.62.
A plot of the relative fit error as a function of the parameters $g$ and $e^*/e$ is shown in Fig. \ref{TDep83}.c. 
For the best fit, $\chi^2=1.20$ is found here, thus indicating only a small systematic disagreement.
Marginalization of $\sigma$, $A$ and $g_\infty$ reveals that the maximum probability is not exactly coinciding with the minimum relative fit error, but slightly shifted to $e^*/e$ = 0.23, $g$ = 0.65 (Fig. \ref{TDep83}.d).
Parameters of the candidate wave functions for $\nu$ = 8/3 are cited in Table \ref{TabelleParameters} and are indicated as (green) circles {(Abelian modes) or (green) dots (non-Abelian modes)} in Fig. \ref{TDep83}.c,d.
All candidate states furthermore exhibit Abelian 2$e$/3 QP excitations with $g$=2/3, which were not observed in the experiment.
{Apart from a particle-hole conjugate Laughlin state ($\overline{\mathrm{L}_{1/3}}$), two types of Bonderson-Slingerland states (BS$_{2/3}$ and $\overline{\mathrm{BS_{1/3}^\psi }}$) and a four-clustered Read-Rezayi state are possible candidates. The Bonderson-Slingerland states are constructed hierarchically over a Moore-Read Pfaffian state. This construction allows to produce the most important filling factors in the second Landau level \cite{bonderson_fractional_2008}. In the four-clustered Read-Rezayi state (RR$_\mathrm{k=4}$), clusters of $k$ anyons are expected to form effective bosons and to condense in a liquid of filling factor $\nu=k/(k+2)$ \cite{read_beyond_1999,stern_anyons_2008}. }
The BS$_{2/3}$ and $\overline{\mathrm{BS_{1/3}^\psi }}$ states support two $e/3$ edge modes with $g$ = 2/3 and 7/24 (BS$_{2/3}$) and $g$ = 2/3 and 13/24 ($\overline{\mathrm{BS_{1/3}^\psi }}$). 
Due to the $g_\mathrm{tun} \propto T^{2g-2}$ temperature dependence, we expect to probe mainly the smallest $g$ of the edge modes. From Fig. \ref{TDep83}.c,d we can see that the RR$_\mathrm{k=4}$ state and the non-Abelian edge modes of the BS$_{2/3}$ state are not in agreement with our measurements. 
The fit parameters are closest to the particle-hole conjugate Laughlin state ($\overline{\mathrm{L}_{1/3}}$), which fits much better than the non-Abelian edge modes of the $\overline{\mathrm{BS_{1/3}^\psi }}$ state. The experimental $g_\mathrm{tun}$ is qualitatively well reproduced by the $\overline{\mathrm{L}_{1/3}}$ parameters (see Supplementary Information). Quantitatively however, none of the candidate states lies within statistical error bars, also in this case. Similar to the previous case, "best fit" parameters can be found that lead to only a statistic deviation of theory and experiment. Again, the deviation of proposed parameter pairs and ``best fit" parameters indicates a systematic deviation.

\renewcommand{\arraystretch}{1.3}
\begin{table*}
\begin{footnotesize}
\begin{minipage}[t]{0.4\linewidth}
\begin{tabular}{cccc}
\multicolumn{4}{c}{$\nu$=5/2}  \\
\toprule
 \multicolumn{4}{c}{Theory \cite{bishara_interferometric_2009,yang_influence_2013,overbosch_phase_2008}} \\
 State 					& $e^*/e$				& $g$ 	& n-A?	\\ \toprule
 $K$=8 			 			& 		1/4				 	& 1/8	 											& No\\ \cline{1-4}
 MR Pf 						& 		1/4					&  1/4													& 	 Yes\\\cline{1-4}
(3,3,1) 						&  	1/4					& 3/8												& No\\\cline{1-4}
 { (1,1,3) }						&  	1/4					& $\approx$ 3/8 \footnote{in a gate defined geometry \cite{yang_experimental_2014}}											& No\\\cline{1-4}
$\overline{\mathrm{Pf}}$					&		1/4					& 	1/2											& Yes\\\cline{1-4}
 SU(2)$_2$						&  	1/4					& 	1/2											& 					Yes\\\cline{1-4}
  $\overline{\mathrm{(3,3,1)}}$						&  	1/4					& 	5/8											& No	\\\cline{1-4}
 $\overline{\mathrm{SU_2(2)}}$						&  	1/4				& 	3/4											& Yes	 \\\cline{1-4}
  {Majorana-gapped}						&  	1/4					& 	1/2											& 	Yes \\\cline{2-4}
{edge-rec. Pf} 				&  	1/2					& 	1/2											& No	\\\cline{1-4}
 { Majorana-gapped}					&  	1/4					& 	0.55-0.75											& Yes	\\\cline{2-4}
{$\overline{\mathrm{Pf}}$}						&  	1/2					& 	0.5-0.7											& No	\\\toprule
 \multicolumn{4}{c}{Experiment} \\
Config.&$e^*/e$ & $g$\\ \toprule
 \textbf{I.}	& 0.18 & 0.32	\\ \cline{1-4}
 \textbf{II.}& 0.25 & 0.42		\\\cline{1-4}
  \textbf{III.}	& 1/4 & 0.42	\\\cline{1-4}
 \textbf{IV.}		& 0.15 - 0.21 &		0.24 - 0.32\\\cline{1-4}
   \textbf{V.}	& 1/4 &	0.37 \\\toprule
\end{tabular}
\end{minipage}\hfill
\begin{minipage}[c]{0.3\linewidth}
\begin{tabular}{cccc}
\multicolumn{4}{c}{$\nu$=8/3}  \\
\toprule
 \multicolumn{4}{c}{Theory \cite{bishara_interferometric_2009}} \\
 State 					& $e^*/e$				& $g$ 	& n-A?	\\ \toprule	 	
 $\overline{\mathrm{L_{1/3}}}$ 			 			& 		1/3				 	& 2/3	 											& No\\ \cline{1-4}
 \multirow{2}*{BS$_{2/3}$} 						& 		1/3					&  7/24													& 	 Yes		\\\cline{2-4}
						& 		1/3					&  2/3													& 	 No		\\\cline{1-4}
  \multirow{2}*{$\overline{\mathrm{BS_{1/3}^\psi }}$} 						&  	1/3					& 13/24												& Yes		\\\cline{2-4}
   						&  	1/3					& 2/3												& No\\\cline{1-4}
 	RR$_{k=4}$				&		1/6					& 	1/6											& Yes		\\\toprule	
  \multicolumn{4}{c}{Experiment} \\
Config.&$e^*/e$ & $g$\\ \toprule	 	 	
  \textbf{I.}	& 0.22 & 0.62	\\ \cline{1-4}
 	\textbf{II.}		& 0.21 - 0.25 & 0.55 - 0.72			\\\cline{1-4}
		\textbf{III.}	& 1/3 & 0.67 - 0.82			\\\cline{1-4}
  	\textbf{IV.}		& 0.19 - 0.28 & 0.62 - 0.84	\\\cline{1-4}
   		\textbf{V.}	& 1/3 & 0.76 - 0.88\\\toprule	
\end{tabular}
\end{minipage}\hfill
\begin{minipage}[c]{0.3\linewidth}
\begin{tabular}{cccc}
\multicolumn{4}{c}{$\nu$=7/3}  \\
\toprule
 \multicolumn{4}{c}{Theory \cite{bishara_interferometric_2009}} \\
 State 					& $e^*/e$				& $g$ 	& n-A?	\\ \toprule
 $\mathrm{L_{1/3}}$ 			 			& 		1/3				 	& 1/3	 											& No\\ \cline{1-4}
 \multirow{2}*{$\overline{\mathrm{BS}_{2/3}}$} 						& 		1/3					&  23/24													& 	 Yes\\\cline{2-4}
 						& 		1/3					&  1/3													& 	 No	\\\cline{1-4}
  \multirow{2}*{$\mathrm{BS_{1/3}^\psi }$} 						&  	1/3					& 17/24												& Yes\\\cline{2-4}
  						&  	1/3					& 1/3												& No\\\cline{1-4}
	$\overline{\mathrm{RR}}_{k=4}$				&		1/6					& 	1/3											& Yes	\\\toprule	
  \multicolumn{4}{c}{Experiment} \\
Config.&$e^*/e$ & $g$\\ \toprule
	\textbf{I.}	& - & -							\\ \cline{1-4}
	\textbf{II.}	& 0.21 - 0.29 & 0.34 - 0.45		\\\cline{1-4}
 		\textbf{III.}	& 1/3 & 0.47	\\\cline{1-4}
 	\textbf{IV.} &0.28 & 0.49		\\\cline{1-4}
   	\textbf{V.}	& 1/3 &	 0.54			\\\toprule	
\end{tabular}
\end{minipage}
\caption{Overview of the theoretically proposed parameter pairs $g$ and $e^*/e$ for different states ('n-A': non-Abelian; taken from Refs. \onlinecite{bishara_interferometric_2009,yang_influence_2013,overbosch_phase_2008}) and our results for different measurement configurations. Only edge modes with the lowest QP charge are quoted, as they dominate tunneling in our experiment. [Config. \textbf{I}.: $B$ and $V_\mathrm{QPC}$ constant (see Figs. \ref{TDep},\ref{TDep83}), \textbf{II}.: $V_\mathrm{QPC}$ varied (Fig. \ref{QPC_Parameters}), \textbf{III}.: $V_\mathrm{QPC}$ varied, $e$* fixed to $e$/4 or $e$/3 (Fig. \ref{QPC_Parameters}), \textbf{IV}.: $B$ varied (Fig. \ref{B_Bias}), \textbf{V}.: $B$ varied, $e$* fixed to $e$/4 or $e$/3 (Fig. \ref{B_Bias})]}
\label{TabelleParameters}

\end{footnotesize}
\end{table*}

\subsection{Effect of varying the coupling via the magnetic field}
\label{Bvar}
\begin{figure}
	\begin{center}
	\includegraphics[width=8.6cm]{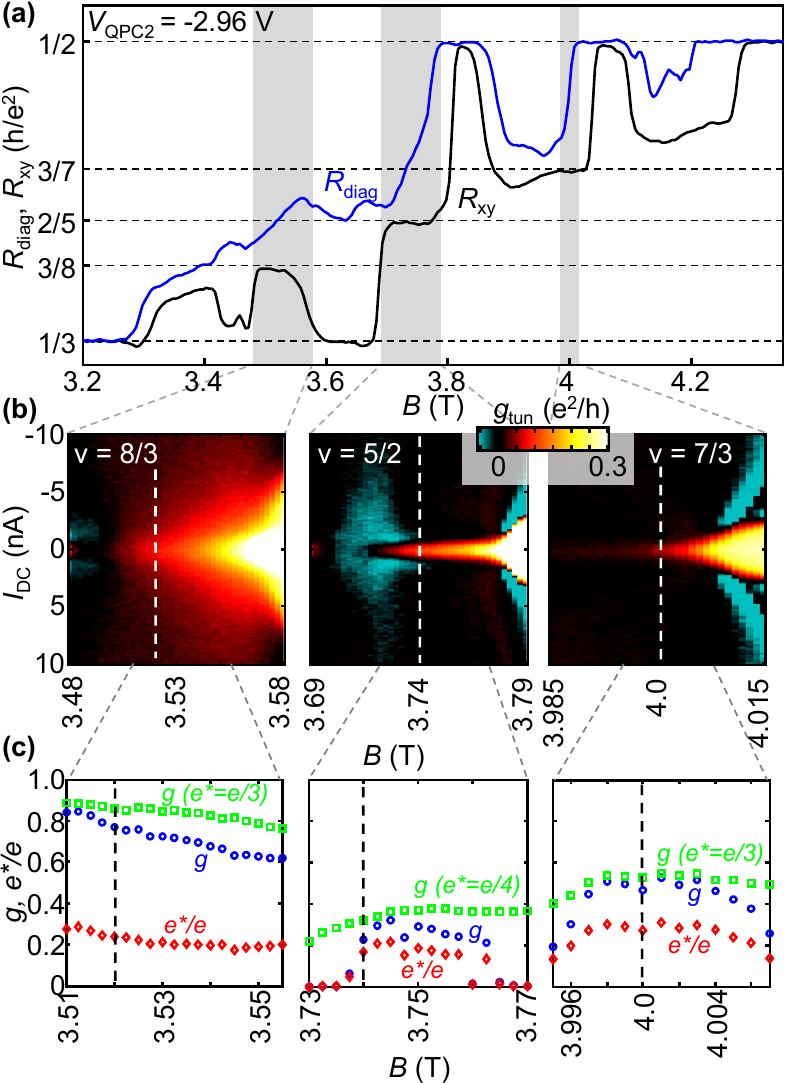}
	\end{center}
	\caption{(Color online) \textbf{a,b:} $B$-field and DC bias dependence of the tunneling conductance near $\nu$=8/3, 5/2 and 7/3. \textbf{c:} $B$-field dependence of fitting parameters $e^*/e$ [(red) diamonds], $g$ [(blue) circles] and $g$ for $e^*/e$ fixed to 1/3 or 1/4 [(green) squares]. The vertical dashed lines in (\textbf{b}) and (\textbf{c}) indicate the center of the bulk filling factor plateaus. }
	\label{B_Bias}
	\end{figure}

The discussed measurements leave the question of how the extracted parameters $e^*/e$ and $g$ depend on the strength of QP tunneling. 
To investigate this, the QPC transmission has been varied by changing the $B$-field. The tunneling conductance $g_\mathrm{tun}$ is shown as a function of DC bias and $B$-field in Fig. \ref{B_Bias}.b.
The $B$-field has been varied in a small window around the bulk filling factors 8/3, 5/2 and 7/3 (gray shaded areas in Fig. \ref{B_Bias}.a). 
Here, QPC2 instead of QPC1 was used in a different cool-down and $V_\mathrm{QPC2}$ was fixed at -2.96 V. As the magnetic field strength is increased, backscattering and hence $g_\mathrm{tun}$ continuously increase. 
At the same time we move out of the $B$-field range where the FQH states are fully gapped in the bulk. Hence the interpretation of the QP backscattering peak only makes sense in narrow $B$-field regions around the bulk filling factors observed in $R_\mathrm{xy}$.
At the high magnetic field end of the graphs, reentrant integer quantum Hall (RIQH) states enter and dominate the temperature dependence of the conductance, resulting in a zero-bias peak with increased FWHM. Due to the complicated and dominant temperature-dependence of the RIQH states \cite{deng_collective_2012}, a qualitative description via Eq. \ref{TunnelFormula} breaks down as soon as they contribute to the conductance.
Away from these states, the FWHM of $g_\mathrm{tun}$ is constant over a wide $B$-field interval (see Supplementary Information). 
The parameters $g$ and $e^*/e$ are extracted from temperature dependent measurements of Fig. \ref{B_Bias}.b. They are shown in Fig. \ref{B_Bias}.c for the $B$-field interval in which the peak FWHM is constant.
Fits of $g_\mathrm{tun}$ (Eq. \ref{TunnelFormula}) yield $g$ [(blue) circles] and $e^*/e$ [(red) diamonds]. When $e^*/e$ is not used as a fitting parameter but fixed at 1/3 ($\nu$ = 7/3, 8/3) or 1/4 ($\nu$ = 5/2), another set of $g$ [(green) squares] is obtained.

\subsubsection{\texorpdfstring{$\nu$}{\textbackslash nu}=8/3}
For $\nu$ = 8/3 (Fig. \ref{B_Bias}.b,c, left column), a continous decrease of $g$ is observed for an increasing $B$-field. When the $B$-field moves away from the bulk 8/3 plateau (at $B$ $>$ 3.56 T), the zero-bias peak shape changes (similar to Fig. \ref{ClosingQPC}.a) and hence is not well described by weak tunneling theory any more. In the $B$-field range where the peak FWHM is constant and no flat peak is observed (Fig. \ref{B_Bias}.c), $g$ varies from 0.82 to 0.62 with $g$ = 0.77 in the center of the bulk $\nu$ = 8/3 plateau (indicated by dashed line, Fig. \ref{B_Bias}.c). At the same time, $e^*/e$ decreases from 0.28 to 0.20 where it saturates. If $e^*/e$ is fixed to 1/3, we find slightly higher values for $g$ in the range 0.88 - 0.76.

\subsubsection{\texorpdfstring{$\nu$}{\textbackslash nu}=5/2}
For $\nu$ = 5/2 (Fig. \ref{B_Bias}.b,c, middle column), a large region of negative differential tunneling conductance $g_\mathrm{tun}$ is observed towards the low-field end of the $\nu$ = 5/2 plateau. The origin of this is not clear. In this case, the undershoots of $g_\mathrm{tun}$ dominate the fit, yielding small values for $g$. Towards the center of the $\nu$ = 5/2 plateau, $g$ and $e^*/e$ take values of $g$ = 0.24 - 0.32 and $e^*/e$ = 0.15 - 0.21. If $e^*/e$ is fixed to 1/4, $g$ saturates at approximately 0.37.

\subsubsection{\texorpdfstring{$\nu$}{\textbackslash nu}=7/3}
For $\nu$ = 7/3 (Fig. \ref{B_Bias}.b,c, right column), a zero-bias peak with constant FWHM is only observed in a very narrow $B$-field window (Fig. \ref{B_Bias}.c). 
Also the amplitude of $g_\mathrm{tun}$ is much smaller than for $\nu$ = 8/3 and 5/2.
At the low-field side of this window, the amplitude of the zero-bias peak is too small for a reliable fit of the data over the whole temperature range. 
At the high-field side, neighboring RIQH states dominate the temperature-dependence of the conductance and broaden the zero-bias peak. 
In between those regimes (where also the bulk plateau center is located, dashed line Fig. \ref{B_Bias}.c), $g$ and $e^*/e$ are approximately constant, with $g$ = 0.49 and $e^*/e$ = 0.28. Fixing $e^*/e$ to 1/3, a plateau value of $g$ $\approx$ 0.54 is obtained.

\subsection{Effect of varying the coupling via the QPC transmission}
\label{QPCvar}
\begin{figure*}
	\begin{center}
	\includegraphics[width=17.2cm]{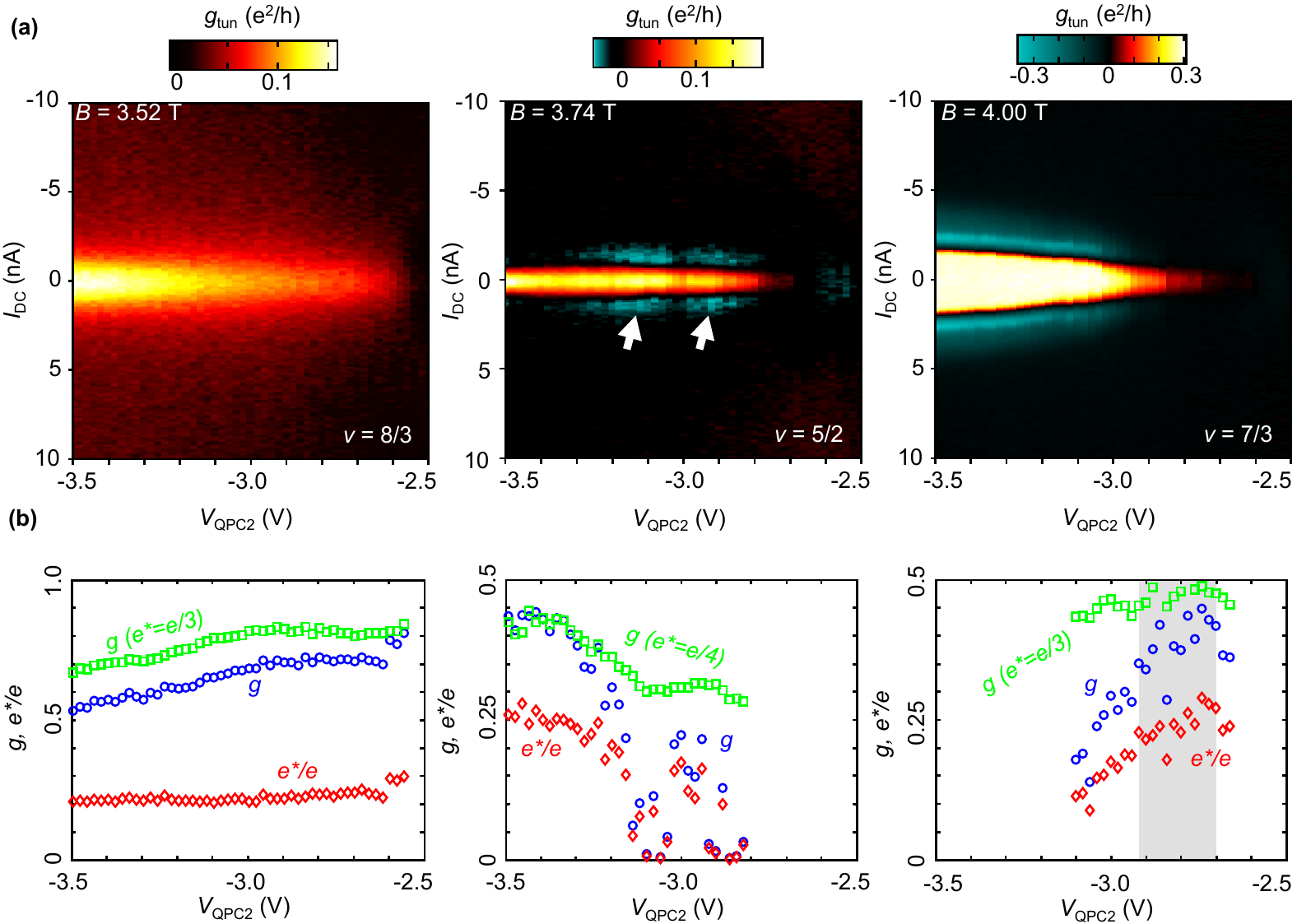}
	\end{center}
	\caption{(Color online) \textbf{a:} Dependence of the tunneling conductance at $\nu$=8/3, 5/2 and 7/3 on the QPC voltage. \textbf{b:} Fitting parameters $e^*/e$ [(red) diamonds], $g$ [(blue) circles] and $g$ for $e^*/e$ fixed to 1/3 or 1/4 [(green) squares] as a function of $V_\mathrm{QPC}$}
	\label{QPC_Parameters}
	\end{figure*}

When the magnetic field is varied, we vary the transmission but might also move out of the gap of the investigated FQH states. Instead, the transmission can be controlled by changing the QPC split-gate voltage while the $B$-field is fixed to the center of the $\nu_\mathrm{bulk}$ = 8/3, 5/2 and 7/3 plateaus.
 When the QPC is closed (Fig. \ref{QPC_Parameters}.a), the amplitude of $g_\mathrm{tun}$ increases. For $\nu$ = 5/2 and 8/3, its FWHM is constant over the whole voltage range, whereas at $\nu$ = 7/3, the FWHM increases due to the neighboring RIQH state. The voltage range in which the FWHM is constant is indicated by the gray shaded area in Fig. \ref{QPC_Parameters}.b.
Parameters $g$ and $e^*/e$, extracted from the temperature dependence of Fig. \ref{QPC_Parameters}.a, are shown in Fig. \ref{QPC_Parameters}.b. 
	
\subsubsection{\texorpdfstring{$\nu$}{\textbackslash nu}=8/3}
At $\nu$ = 8/3 (Fig. \ref{QPC_Parameters}, left column), $e^*/e$ is approximately constant at 0.21 - 0.25  over the whole gate voltage range. For $g$, constant values of approx. 0.72 are found for small $g_\mathrm{tun}$, which start to decrease at $V_\mathrm{QPC}~\approx~-3.0$ V down to $g$ = 0.55. If $e^*/e$ is fixed to 1/3, a similar evolution of $g$, with slightly higher values ($g$ = 0.67 - 0.82) is found.
 
\subsubsection{\texorpdfstring{$\nu$}{\textbackslash nu}=5/2}
Here (Fig. \ref{QPC_Parameters}, middle column), the FWHM of $g_\mathrm{tun}$ is constant over the whole voltage range. At $V_\mathrm{QPC}~\approx$ -3.15 V and $V_\mathrm{QPC}~\approx$ -2.9 V (marked by white arrows), the $g_\mathrm{tun}$ peak is locally enhanced and neighbored by negative differential conductance undershoots.  This behavior could be caused by resonant tunneling through a localization in the QPC. Here, the $g_\mathrm{tun}$ undershoots dominate the fit, yielding small values for $e^*/e$ and $g$. Towards $V_\mathrm{QPC}$ = -3.5 V, $e^*/e$ and $g$ saturate at 0.25 and 0.42 respectively. For $e^*/e$ fixed to 1/4, $g$ varies from approx. 0.28 at the onset of the zero bias peak to approx. 0.42 at $V_\mathrm{QPC}$ = -3.5 V.

\subsubsection{\texorpdfstring{$\nu$}{\textbackslash nu}=7/3}
 For $\nu$ = 7/3 (Fig. \ref{QPC_Parameters}, right column), $e^*/e$ = 0.21  - 0.29 are observed in the narrow region of constant FWHM (shaded gray). At the same time, we find $g$ = 0.34 - 0.45. Towards more negative $V_\mathrm{QPC}$, the RIQH temperature dependence again dominates and a reliable fit is not possible. Fixing $e^*/e$ to 1/3, an approximately constant $g$ = 0.47 is found. 

\subsection{Breakdown of the weak tunneling regime}
\begin{figure}
	\begin{center}
	\includegraphics[width=8cm]{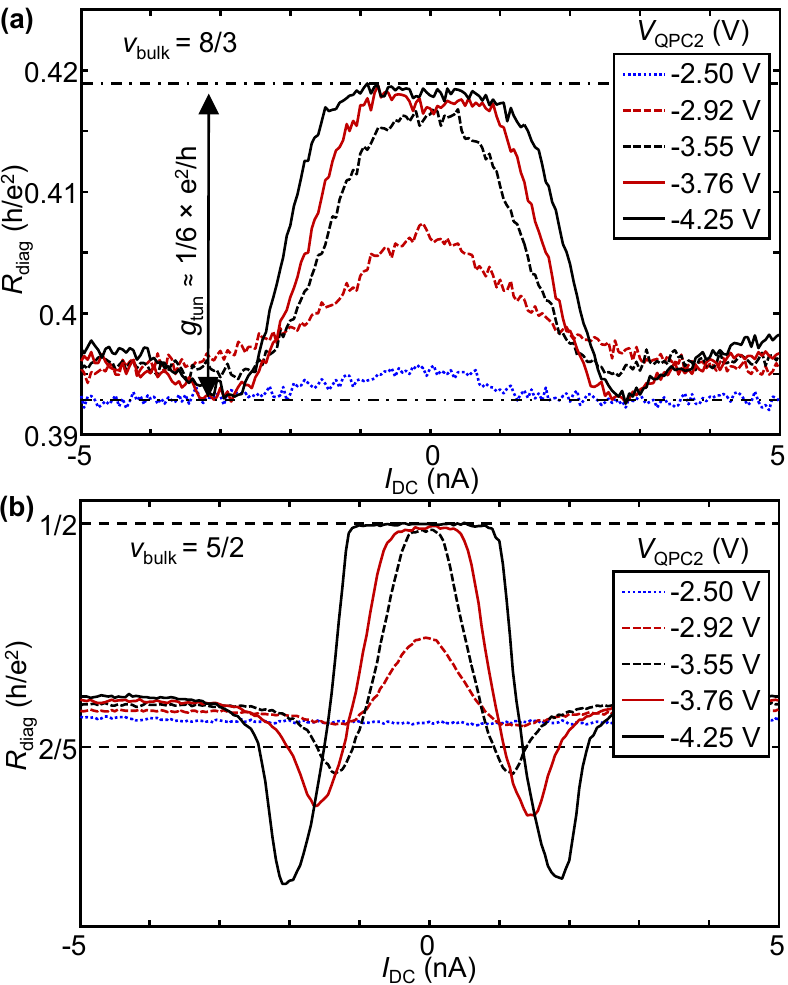}
	\end{center}
	\caption{(Color online) Diagonal resistance across QPC2 (\textbf{a}: $\nu_\mathrm{bulk}$=8/3, \textbf{b}: $\nu_\mathrm{bulk}$=5/2) for different QPC voltages as a function of DC bias. }
	\label{ClosingQPC}
	\end{figure}

As the QPC is pinched off further, a situation can arise in which QPC and bulk have different filling factors. This intermediate tunneling regime has been studied theoretically \cite{fendley_exact_1995} and experimentally \cite{roddaro_interedge_2004,roddaro_quasi-particle_2004} in detail. 
Fig. \ref{ClosingQPC}.a and \ref{ClosingQPC}.b show $R_\mathrm{diag}$ as QPC 2 is biased very negatively at bulk filling factors 8/3 and 5/2.
When QPC 2 is relatively open (meaning the absence of QP tunneling at $V_\mathrm{QPC}$ = -2.5 V), $R_\mathrm{diag}$ is approximately constant at a value slightly higher than expected for the respective bulk filling factor. As the QPC voltage is decreased, a QP tunneling peak at zero DC bias develops. At higher DC biases, $R_\mathrm{diag}$ drops close to the flat background value where it is approximately constant. 
For $\nu_\mathrm{bulk}$ = 8/3, the QP tunneling peak grows, until $R_\mathrm{diag}$ $\approx$ 0.42 $h/e^2$, where $R_\mathrm{diag}$ develops a plateau in $I_\mathrm{DC}$ direction. 
The difference in diagonal resistance between those two values corresponds to $g_\mathrm{tun}\approx 1/6\times e^2/h$ (Fig. \ref{ClosingQPC}.a). It should be noted that this is equal to $(8/3-5/2)\times e^2/h$. 
Thus the situation of Fig. \ref{ClosingQPC}.a might be interpreted as the case where the $\nu=8/3$ edge state is partially reflected from the QPC, leaving a gapped $\nu=5/2$ state within. As the DC bias is increased, the gap is destroyed and the QPC filling approaches 8/3 again.
At a bulk filling factor of 5/2, a transition to a RIQH state is observed (Fig. \ref{ClosingQPC}.b). Here, $R_\mathrm{diag}$ is quantized at exactly $2\times e^2/h$. As the DC bias is increased, strong undershoots in $R_\mathrm{diag}$ are observed. Then $R_\mathrm{diag}$ saturates again at around $R_\mathrm{diag}\approx 2/5\times h/e^2$.

\section{Interpretation and discussion}
\subsection{\texorpdfstring{$\nu$}{\textbackslash nu}=8/3}
For $\nu$ = 8/3, all results (see Table \ref{TabelleParameters} for an overview) favor the proposed parameters $g$ = 2/3 and $e^*/e$ = 1/3. 
The BS$_{2/3}$ and $\overline{\mathrm{BS_{1/3}^\psi }}$ states support additional non-Abelian $e/3$ edge modes with $g$ = 7/24 (BS$_{2/3}$) and $g$ = 13/24 ($\overline{\mathrm{BS_{1/3}^\psi }}$) which should dominate the temperature dependence.
Thus the measurements agree best with an Abelian particle-hole conjugate Laughlin state ($\overline{\mathrm{L_{1/3}}}$), which qualitatively well reproduces our measurements (see Supplementary Information).
In the $V_\mathrm{QPC}$ dependent measurement, a constant $g$ of approximately 0.72 is observed at the onset of QP tunneling. As the tunneling strength increases, $g$ decreases to 0.55. This might either be caused by additional coupling due to a second edge mode, or the breakdown of the weak tunneling assumptions.

\subsection{\texorpdfstring{$\nu$}{\textbackslash nu}=5/2}
Here the interpretation of the results is less clear. 
If $V_\mathrm{QPC}$ and $B$ are kept constant, we find  $e^*/e$ = 0.18 and $g$ = 0.32, close to values reported earlier\cite{radu_quasi-particle_2008,lin_measurements_2012}. For the case of a varying $B$-field, $g$ of 0.29 - 0.32 and $e^*/e$ of 0.19 - 0.21 are observed close to the center of the $R_\mathrm{xy}=2/5\times h/e^2$ plateau. 
If $e^*/e$ is fixed to 1/4, $g$ saturates at around 0.37. This agrees best with the (3,3,1)- and (1,1,3)-states. As the QPC voltage is changed, an evolution of $g$ with a saturation at $g~\approx$ 0.42 is observed. This would also be best described by the parameters of the (3,3,1)- and (1,1,3)-states. 
The origin of the strong modulation of the parameters might be the coupling to a localized state in the QPC, which can substantially influence the transmission \cite{baer_interplay_2014}. However, at the onset of the QP tunneling peak ($V_\mathrm{QPC}\approx-2.9$ V), $g$ $\approx$ 0.30 - 0.31 is found. This parameter lies in-between the expectation for the Moore-Read Pfaffian ($g$ = 0.25) and {the (3,3,1) and (1,1,3)-states} ($g$ = 0.375). 
Overall, our results agree best with the Abelian {(3,3,1) and (1,1,3)-states}, which qualitatively fits the measurements (see Supplementary Information). 
In Ref. \onlinecite{radu_quasi-particle_2008}, the $\overline{\mathrm{Pf}}$ and SU$_2$(2) states were found to be the states with the lowest fit error, whereas Ref. \onlinecite{lin_measurements_2012} also found the (3,3,1)-state to be the best fit. As argued in Ref. \onlinecite{yang_influence_2013}, electron-electron interaction within the edge modifies the effective Coulomb interaction parameter $g$. In this case, both experiments would also be best described by the {(3,3,1) and (1,1,3)-states}.
For a geometry similar to the QPC geometry used by us, the measured $g$ is expected to be enhanced by approximately 0.04 compared to the actual $g$ of the FQH state \cite{yang_influence_2013}. Taking this into account would improve the agreement with the (3,3,1)-state in the case where $e^*/e$ is fixed to 1/4 (Figs. \ref{TDep}.c, \ref{TDep}.d).\\
It should be noted that {the (3,3,1)-state is not compatible with all results obtained by other authors, while the (1,1,3)-state seems to be compatible with all experimental observations. 
{
Numerical diagonalization studies however,} favor the Moore-Read Pfaffian state or its particle-hole conjugate ($\overline{\mathrm{Pf}}$) \cite{morf_transition_1998,rezayi_incompressible_2000,feiguin_density_2008,moller_paired_2008,feiguin_spin_2009,zozulya_entanglement_2009,storni_fractional_2010,moller_neutral_2011,wan_edge_2006,wan_fractional_2008,dimov_spin_2008}. To our knowledge, only the spin-unpolarized version of the (3,3,1)-state has been investigated numerically. The question whether the ground state is better described by the Moore-Read Pfaffian or the Anti-Pfaffian has recently raised interest. Both states cannot be distinguished when particle-hole symmetry is assumed \cite{levin_particle-hole_2007,lee_particle-hole_2007}. The Pfaffian \cite{wojs_landau-level_2010}, as well as the Anti-Pfaffian \cite{rezayi_breaking_2011} have found support in studies employing different numerical approaches.
Finite thickness effects, which might also have to be taken into account for a correct description of the ground state, have been found to stabilize the Pfaffian ground state \cite{peterson_finite-layer_2008,peterson_orbital_2008,biddle_entanglement_2011} and to significantly enhance the overlap with the numerical solution.
}
Recent interference experiments \cite{willett_measurement_2009,willett_alternation_2010,willett_magnetic-field-tuned_2013} might indicate non-Abelian statistics.  Still, this does not rule out the Abelian {(3,3,1) and (1,1,3)-states}, as they might show similar signatures in the interference pattern \cite{stern_interference_2010,yang_experimental_2014}. 
Experiments probing the spin polarization at $\nu$ = 5/2 \cite{zhang_5/2_2010,stern_optical_2010,rhone_rapid_2011,tiemann_unraveling_2012,stern_nmr_2012,wurstbauer_resonant_2013} obtained contradicting results for the polarization. {Recent experiments suggest a spin transition of the $\nu$=5/2 state at very low densities \cite{pan_competing_2014}, similar to $\nu$=8/3 \cite{pan_spin_2012}. Nevertheless,} the (3,3,1)-state exists both in a spin-polarized and spin-unpolarized type \cite{overbosch_phase_2008,yang_influence_2013} with identical Coulomb interaction parameter $g$.
{In contrast, only the spin-polarized version of the $\overline{\mathrm{(3,3,1)}}$ state is allowed for $\nu$=5/2 \cite{yang_influence_2013}. The physical origin of the spin-polarized and spin-unpolarized versions of the (3,3,1)-state is however different. The spin-unpolarized version can be understood as Halperin's bilayer (3,3,1)-state\cite{halperin_theory_1983}, where spin up or down electrons take the function of the two different layers \cite{overbosch_phase_2008,yang_influence_2013}. In contrast, the spin-polarized version arises when charge 2$e$/3 quasiparticles condense on top of a $\nu=1/3$ Laughlin state \cite{yang_influence_2013}. Also the (1,1,3)-state might occur with and without spin polarization \cite{yang_experimental_2014}.}
Shot noise experiments report the observation of a neutral mode for $\nu=5/2$ \cite{bid_observation_2010}.{Such a counterpropagating neutral mode is not expected for the (3,3,1)-state, but for the (1,1,3)-state.}
However, recent experiments \cite{inoue_proliferation_2014} suggest the presence of neutral modes, even for non-particle-hole-conjugate FQH states. Thus, the existence of a neutral mode might not directly allow to draw conclusions about the wave function of the corresponding FQH liquid.\\
As pointed out earlier \cite{jain_5/2_2010,lin_measurements_2012}, these inconsistencies might indicate that the $\nu=5/2$ state might form different wave functions, depending on the physical situation.

\subsection{\texorpdfstring{$\nu$}{\textbackslash nu}=7/3}
For $\nu$ = 7/3, the problem arises that $e^*/e$ = 1/3 and $g$ = 1/3 are proposed for the $\mathrm{L_{1/3}}$ state and the non-Abelian edge modes of the $\overline{\mathrm{BS}_{2/3}}$ and $\mathrm{BS_{1/3}^\psi }$ states. Here, the dominant temperature dependence is expected to be due to the non-Abelian edge modes (smallest $g$), in contrast to the case at $\nu$ = 8/3.
This makes the discrimination of these states in the experiment impossible. Experimentally, we observe $g~>$ 1/3 ($g$ saturates at approximately 0.49 when the $B$-field is varied and $g$ = 0.34 - 0.45 when changing $V_\mathrm{QPC}$), which might stem from a contribution of a second (non-Abelian) edge mode.
The fact that the FQH state at $\nu$ = 8/3 is best described by a particle-hole conjugate Laughlin state ($\overline{\mathrm{L_{1/3}}}$) does not imply that the 7/3 state must be the corresponding non-conjugate partner state ($\mathrm{L_{1/3}}$). As argued in Ref. \onlinecite{bishara_interferometric_2009}, particle-hole symmetry might for example be broken by LL mixing, or other effects.

\subsection{Experimental limitations and origin of systematic errors}
In an ideal system, density is homogeneous and edge states are brought in close proximity by the QPC, until QPs tunnel between two points of the counterpropagating edge states. However, in a realistic system, density is not perfectly homogeneous. 
The coexistence of different FQH states in the bulk and the constriction strongly modifies the system's behavior. For the system studied here, densities of constriction and bulk are sufficiently similar to avoid the coexistence of different FQH states in the second LL (Fig. \ref{BulkDiag}). 
However, density-modulated RIQH states are observed in close proximity to $\nu$=5/2, 7/3 and 8/3. If such states are formed within the constriction, they might strongly modify the temperature scaling of the conductance. 
At $\nu$=5/2 and 8/3, a pronounced zero-bias peak is visible, sufficiently far away from the parameter ranges where a contribution of the RIQH states to the conductance becomes visible (Figs. \ref{B_Bias}.b, \ref{QPC_Parameters}.a). 
Thus we here expect a negligible influence of the density modulated phases. 
However, for $\nu$=7/3, a zero-bias peak is only visible in close proximity to the parameter ranges where the neighboring RIQH state clearly dominates the conductance (Figs. \ref{B_Bias}.b, \ref{QPC_Parameters}.a). 
Although tunneling parameters have been extracted in the regions where the FWHM of the zero-bias peak is constant, a contribution of the neighboring RIQH state cannot be fully excluded.

Another question is the validity of the weak tunneling assumption. In the second LL, the FQH states contribute $G=2\frac{e^2}{h} + \delta G$ to the conductance. For the weak tunneling approximation to hold, $g_\mathrm{tun} \ll \delta G$ is required (if edge reconstruction is present, additional complication might occur). 
At $\nu=8/3$ ($g_\mathrm{tun}\approx0.1\times e^2/h$, $\delta G=2/3\times e^2/h$) and $\nu=7/3$ ($g_\mathrm{tun}\approx0.05\times e^2/h$, $\delta G=1/3\times e^2/h$), this condition is well satisfied within the experimental possibilities.
At $\nu$ = 5/2, we have $\delta G=0.5\times e^2/h$. As the temperature is lowered, $g_\mathrm{tun}$ increases from $g_\mathrm{tun}<0.05\times e^2/h$ to $g_\mathrm{tun}\approx0.15\times e^2/h$. 
Over the whole range, the amplitude of $g_\mathrm{tun}$ is well described by a power law $T^{2g-2}$. When crossing from the weak tunneling regime to the strong tunneling regime, a continuous change of the Coulomb interaction parameter $g$ is expected \cite{kane_edge-state_2007}. 
This is not observed at $\nu$=5/2. Thus we conclude that also in this case we are in, or close to the weak tunneling regime.

Other effects that might cause a systematic measurement error are for example a drift of the QPC transmission and errors in the temperature measurement. However, both of these effects are expected to have a small influence and cannot account for the systematic deviation between measurements and theoretically predicted parameters (Figs. \ref{TDep}.c,d, \ref{TDep83}.c,d).

Furthermore, the tunneling conductance $g_\mathrm{tun}\approx (R_\mathrm{diag}-R_\mathrm{xy})/R_\mathrm{xy}^2$ is an approximation that is only valid in the weak tunneling regime when $R_\mathrm{diag}\approx R_\mathrm{xy}$. For extracting the bias dependence of $g_\mathrm{tun}$, we have assumed that the current reflected at the QPC is much smaller than the current transmitted. These approximations are expected to give an error less than approximately 5$\%$ for $g_\mathrm{tun}$.

\section{Conclusion}
In conclusion, we have measured weak quasiparticle tunneling across a QPC at $\nu$= 8/3, 5/2 and 7/3.
Comparison with theory allowed the extraction of tunneling parameters and comparison with proposed wave functions for these states. A summary of theoretical predictions \cite{bishara_interferometric_2009,yang_influence_2013,overbosch_phase_2008} and of our findings can be found in Table \ref{TabelleParameters}.
Quantitatively, none of the proposed wave functions for $\nu$ = 5/2, 7/3 and 8/3 lies within the statistical error. Qualitatively, the $\nu$ = 5/2 state is well described by an Abelian (3,3,1) or {(1,1,3)-state}. 
However, other experimental findings pose the question of whether the $\nu$ = 5/2 state can manifest in different wave functions, depending on the physical situation. Furthermore we show, that the QP tunneling strength has an impact on extracted tunneling parameters, especially for $\nu$ = 5/2.
For $\nu$ = 8/3, an ordinary particle-hole conjugate Laughlin state reproduces our data best, while proposed non-Abelian edge modes are much less likely. At $\nu$ = 7/3, extracted values for $g$ are not in agreement with the predicted parameters for non-Abelian edge modes. However, the observed parameters $g$ are higher than expected for an ordinary Laughlin state or other Abelian edge modes, which might indicate the presence of several edge modes, in which case identical parameters for different edge modes make a discrimination of the wave functions for $\nu$ = 7/3 impossible.

\section{Acknowledgments}
We gratefully acknowledge discussions with \mbox{R. H. Morf}, \mbox{S. Hennel} and \mbox{C. M. Marcus}. We acknowledge the support of the ETH FIRST laboratory and financial support of the
Swiss Science Foundation (Schweizerischer Nationalfonds, NCCR 'Quantum Science and Technology').

%\section{References}
\bibliographystyle{apsrev4-1}
\bibliography{Bibliothek}

\onecolumngrid
\newpage
\section*{{\LARGE Supplementary information}}
\renewcommand{\thefigure}{S.\arabic{figure}} 
\setcounter{figure}{0}
\label{SuppInf}

\section{FWHM of tunneling peaks}

\begin{figure}[h!]
	\begin{center}
	\includegraphics[width=13cm]{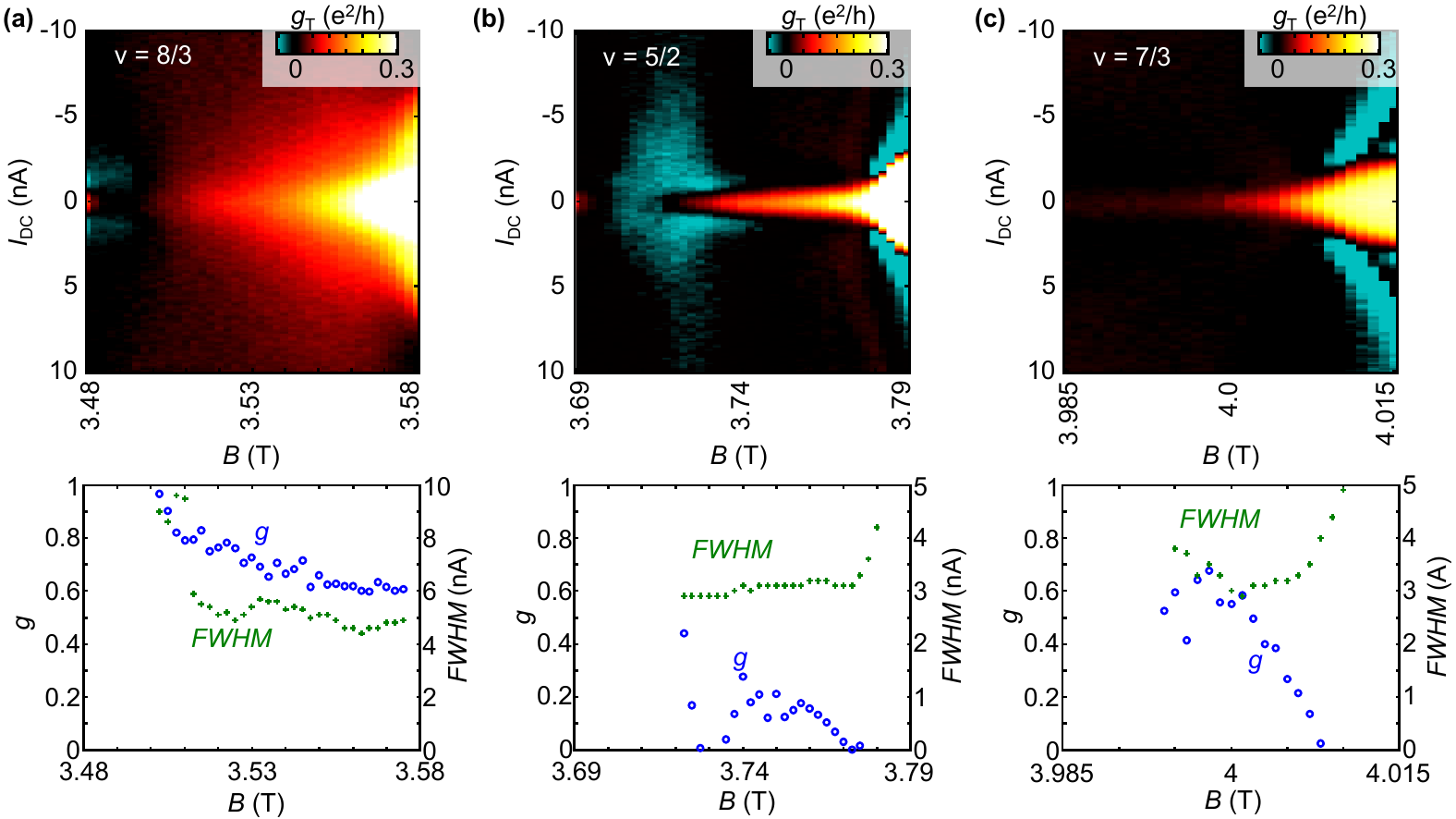}
	\end{center}
	\caption{FWHM of QP tunneling peaks (green crosses) as a function of the $B$-field. Parameters $g$, extracted only from the $I_\mathrm{DC}=0$ scaling of $g_\mathrm{tun}$ are shown as blue circles.}
	\label{FWHM_B}
	\end{figure}

\begin{figure}[h!]
	\begin{center}
	\includegraphics[width=13cm]{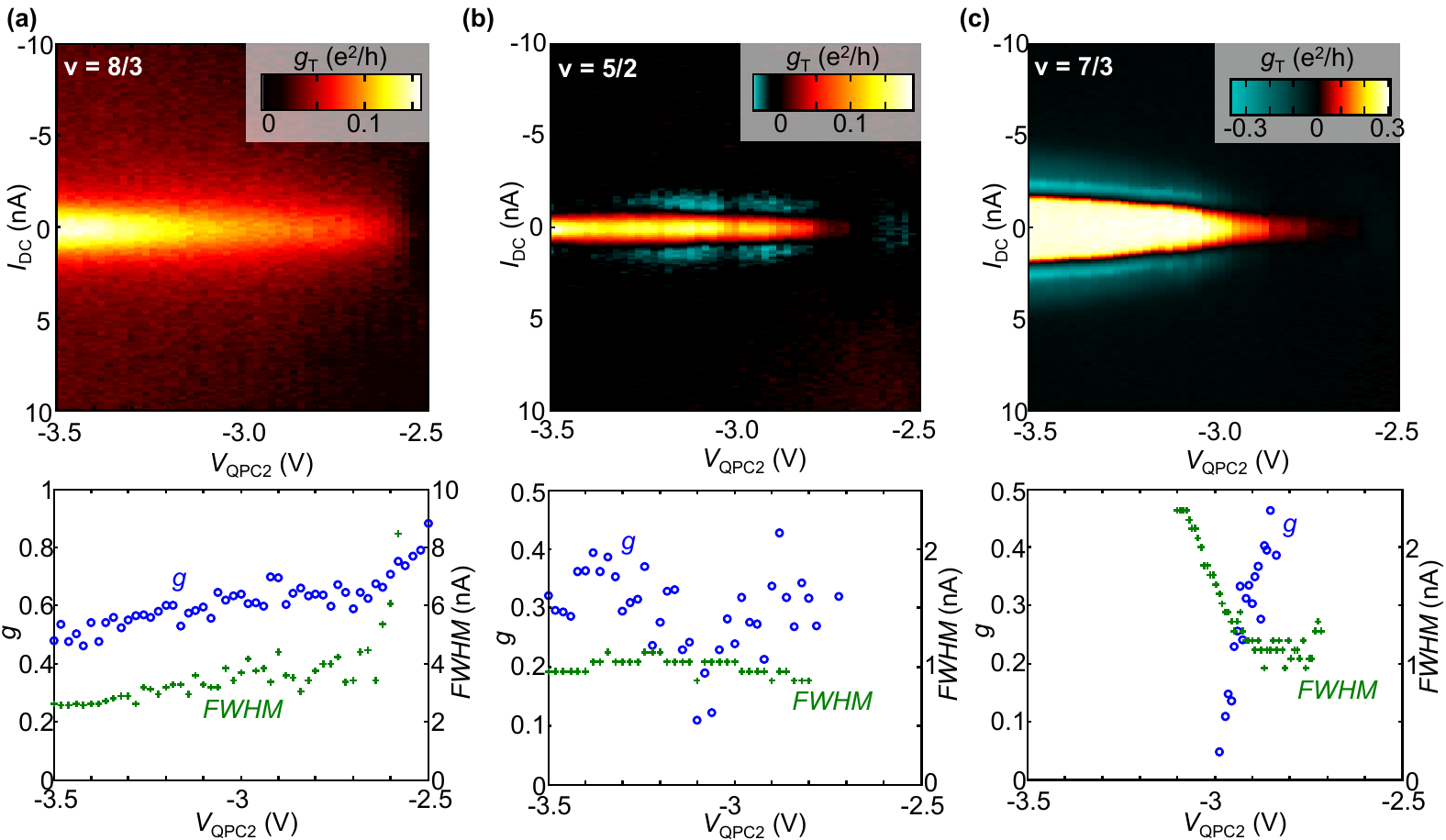}
	\end{center}
	\caption{FWHM of QP tunneling peaks (green crosses) as a function of $V_\mathrm{QPC}$. Parameters $g$, extracted only from the $I_\mathrm{DC}=0$ scaling of $g_\mathrm{tun}$ are shown as blue circles.}
	\label{FWHM_QPC}
	\end{figure}

Figs. \ref{FWHM_B} and \ref{FWHM_QPC} show the  FWHM of the zero-bias peaks at $\nu$=8/3, 5/2 and 7/3. For $\nu$ = 7/3, only a small parameter region with constant FWHM is found. As soon as a neighboring RIQH state contributes to the conductance, the peak FWHM is drastically increased (Figs. \ref{FWHM_B}.b,c and \ref{FWHM_QPC}.c).
Parameters $g$ which are indicated in Figs. \ref{FWHM_B} and \ref{FWHM_QPC} are extracted solely from the $g_\mathrm{tun}\vert_{I_\mathrm{DC}=0}\propto T^{2g-2}$ temperature scaling, in contrast to fits of the whole expression (Eq. 1) used in the main manuscript (Figs. 5, 6). Resulting values for $g$ are similar for both methods.

\newpage

\section{Fits for proposed parameter pairs - \texorpdfstring{$\nu$}{\textbackslash nu} = 5/2}
A qualitative evaluation of the agreement with proposed wave functions can be performed by fixing the parameters $g$ and $e^*/e$ to parameters proposed for different states (see Table I) and fitting $A$ and $g_\infty$. For $\nu$ = 5/2, $g$ = 0.375 and $e^*/e$ = 0.25, corresponding to the (3,3,1)-state, produce a (qualitatively) good agreement with measurement and calculation.

	\begin{figure}[h!]
	\begin{center}
	$\nu$ = 5/2, $g$ = 0.5, $e^*/e$ = 0.25:\\
	\vspace{-0.1cm}
	\includegraphics[width=10cm]{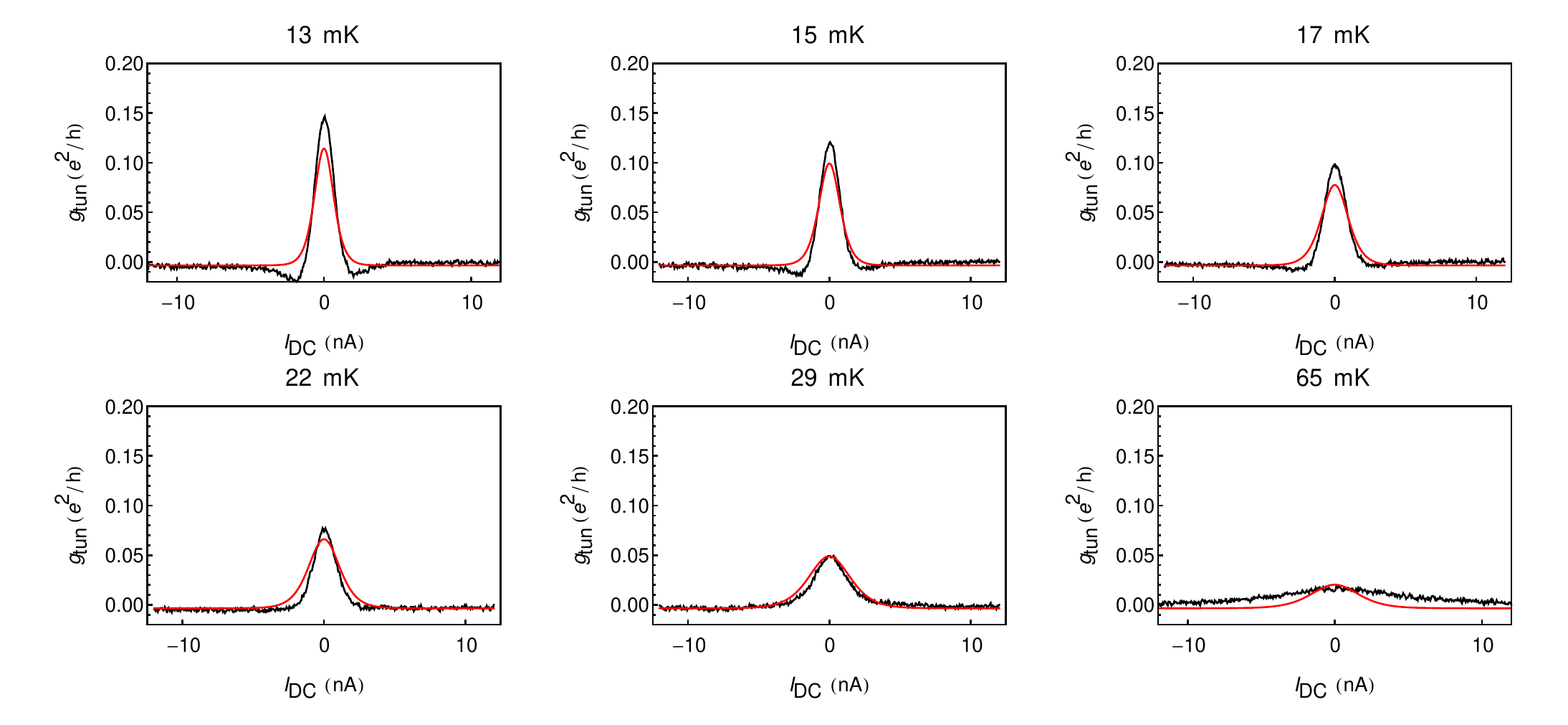}\\
	$\nu$ = 5/2, $g$ = 0.375, $e^*/e$ = 0.25:\\
		\vspace{-0.1cm}
	\includegraphics[width=10cm]{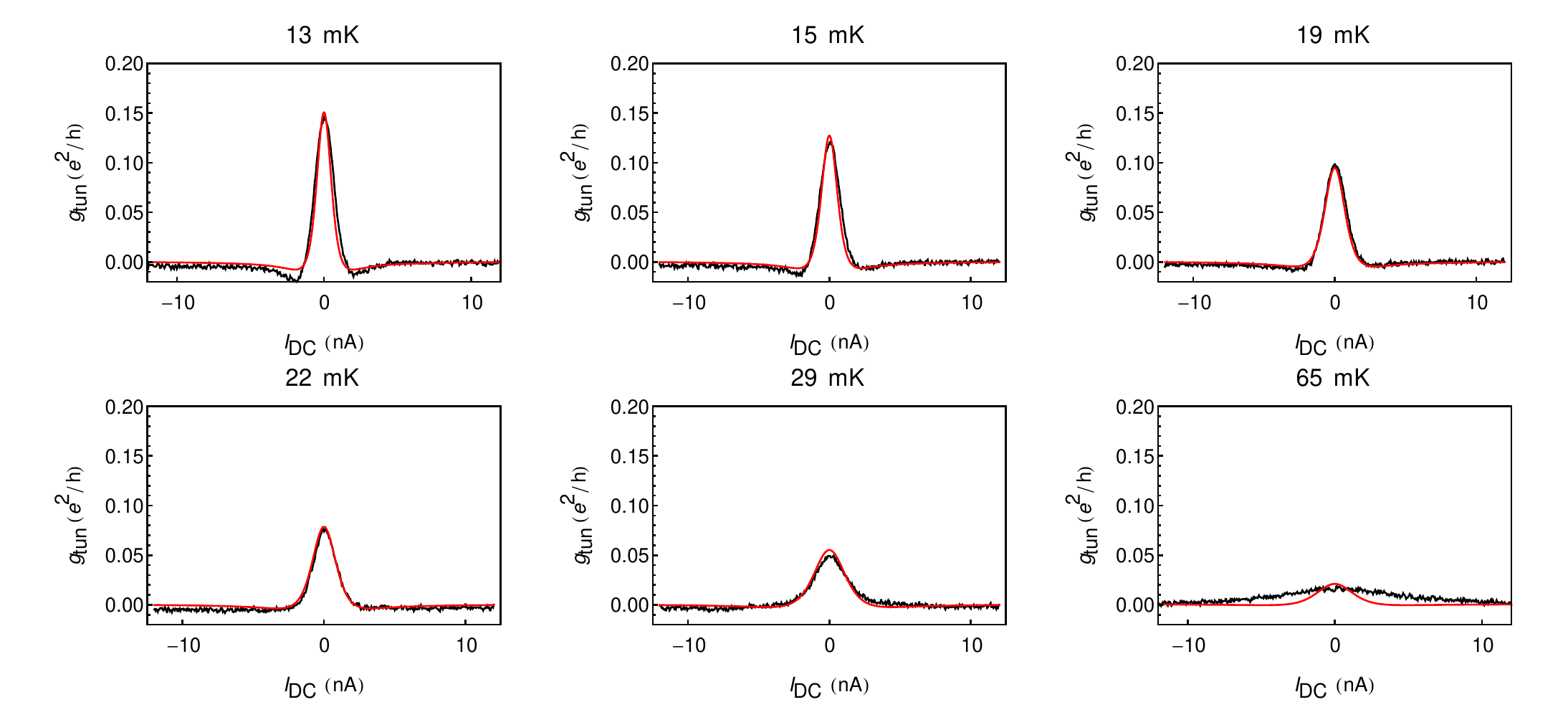}\\
		$\nu$ = 5/2,  $g$ = 0.25, $e^*/e$ = 0.25:\\
			\vspace{-0.1cm}
	\includegraphics[width=10cm]{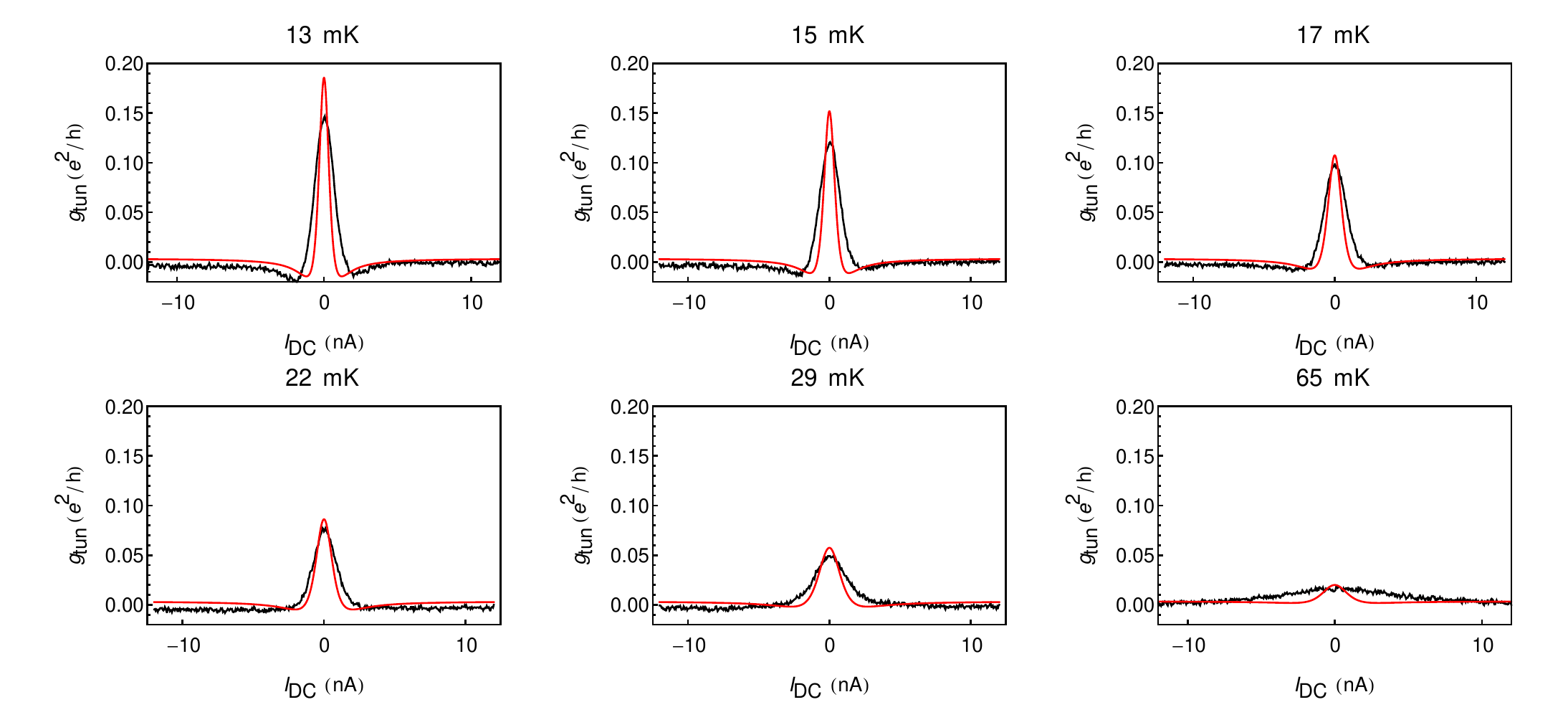}\\
		$\nu$ = 5/2, $g$ = 0.125, $e^*/e$ = 0.25:\\
			\vspace{-0.1cm}
	\includegraphics[width=10cm]{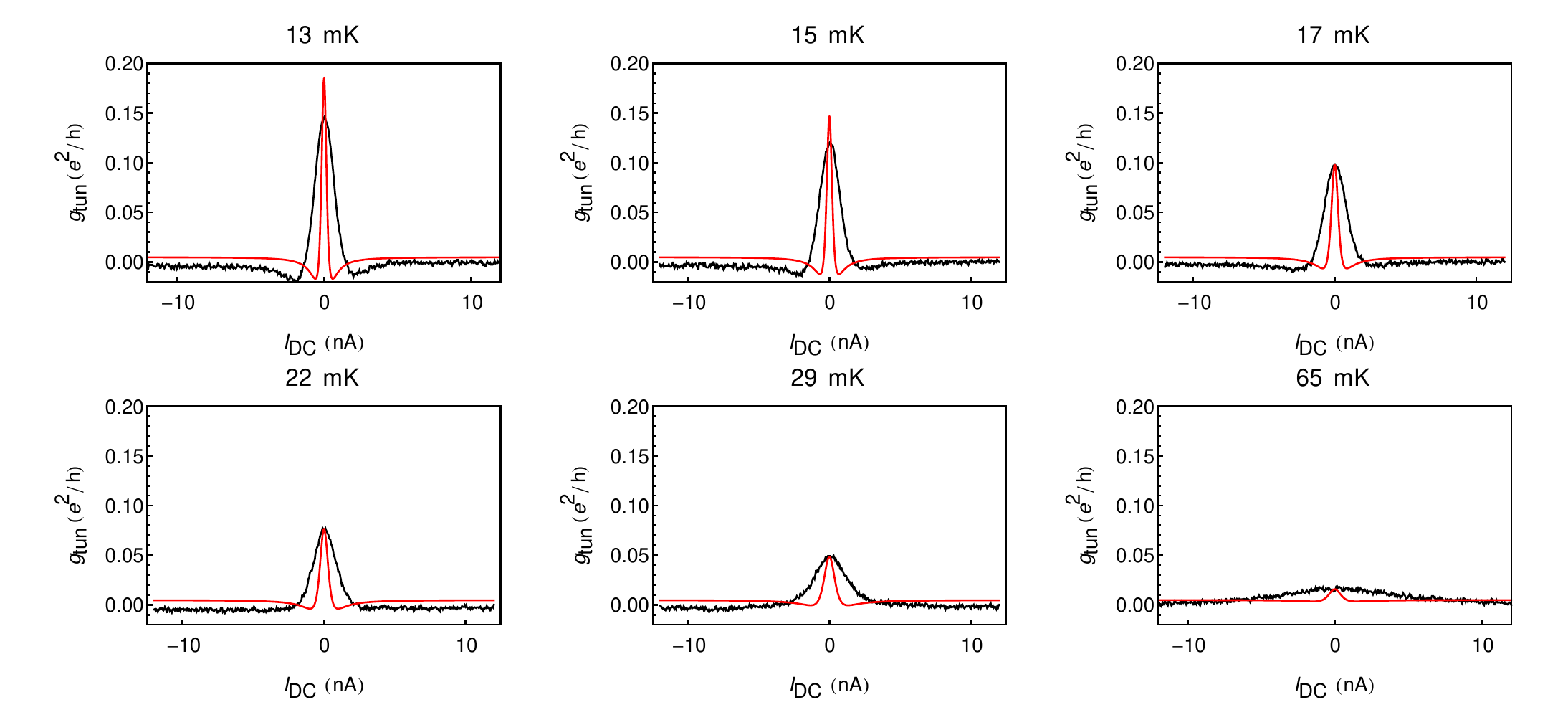}\\
		\end{center}
	\caption{\vspace{-1cm} Comparison between experimental and calculated $g_\mathrm{tun}$ for proposed parameter pairs ($\nu$ = 5/2).}
	\label{Para52}
	\end{figure}

\newpage

\section{Fits for proposed parameter pairs - \texorpdfstring{$\nu$}{\textbackslash nu} = 8/3}
As before, we compare the calculated $g_\mathrm{tun}$ for theoretically proposed parameters with our measurements. For $g$ = 2/3 and $e^*/e$ = 1/3, a (qualitatively) good agreement is found.

	\begin{figure}[h!]
	\begin{center}
	$\nu$ = 8/3, $g$ = 2/3, $e^*/e$ = 1/3\\
	\includegraphics[width=10cm]{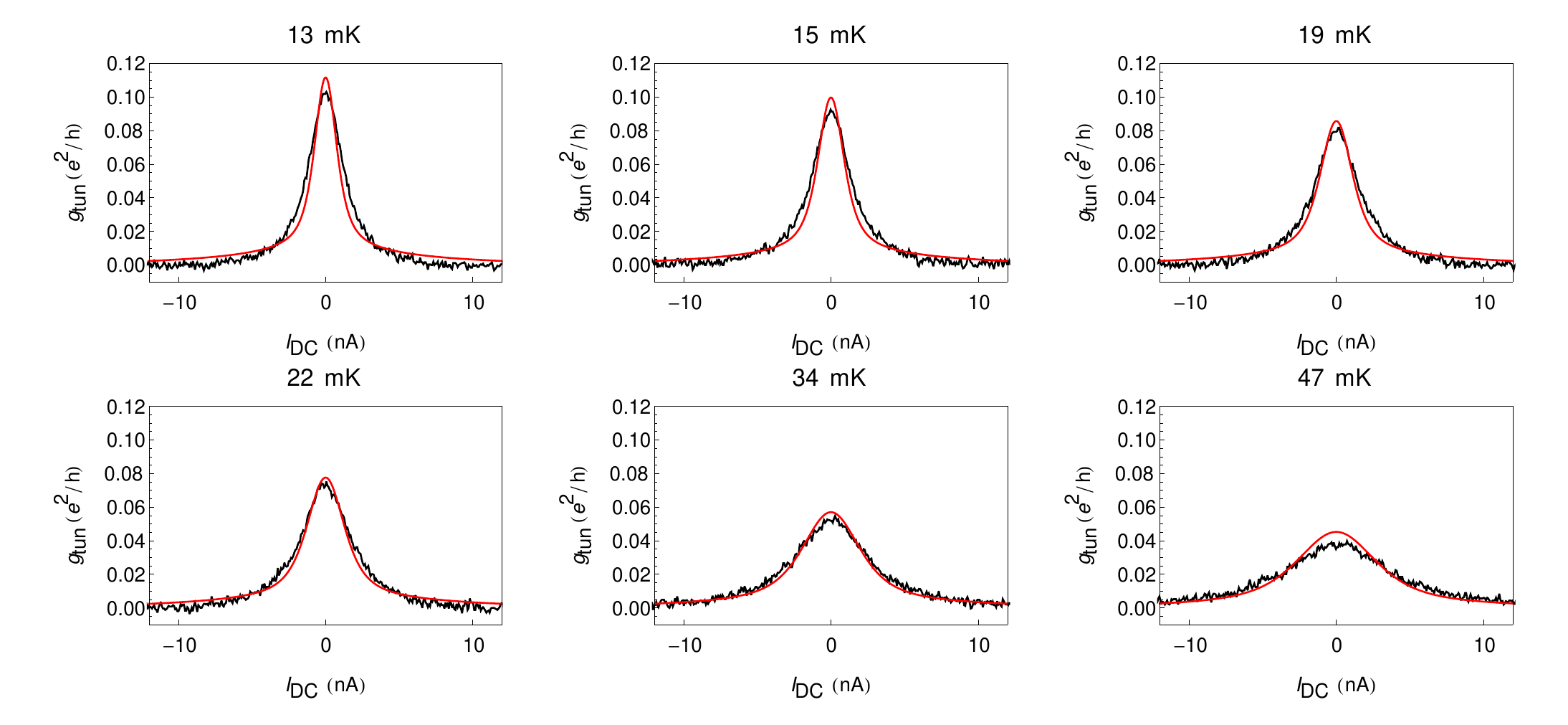}\\
		\vspace{-0.1cm}
	$\nu$ = 8/3, $g$ = 13/24, $e^*/e$ = 1/3\\
	\includegraphics[width=10cm]{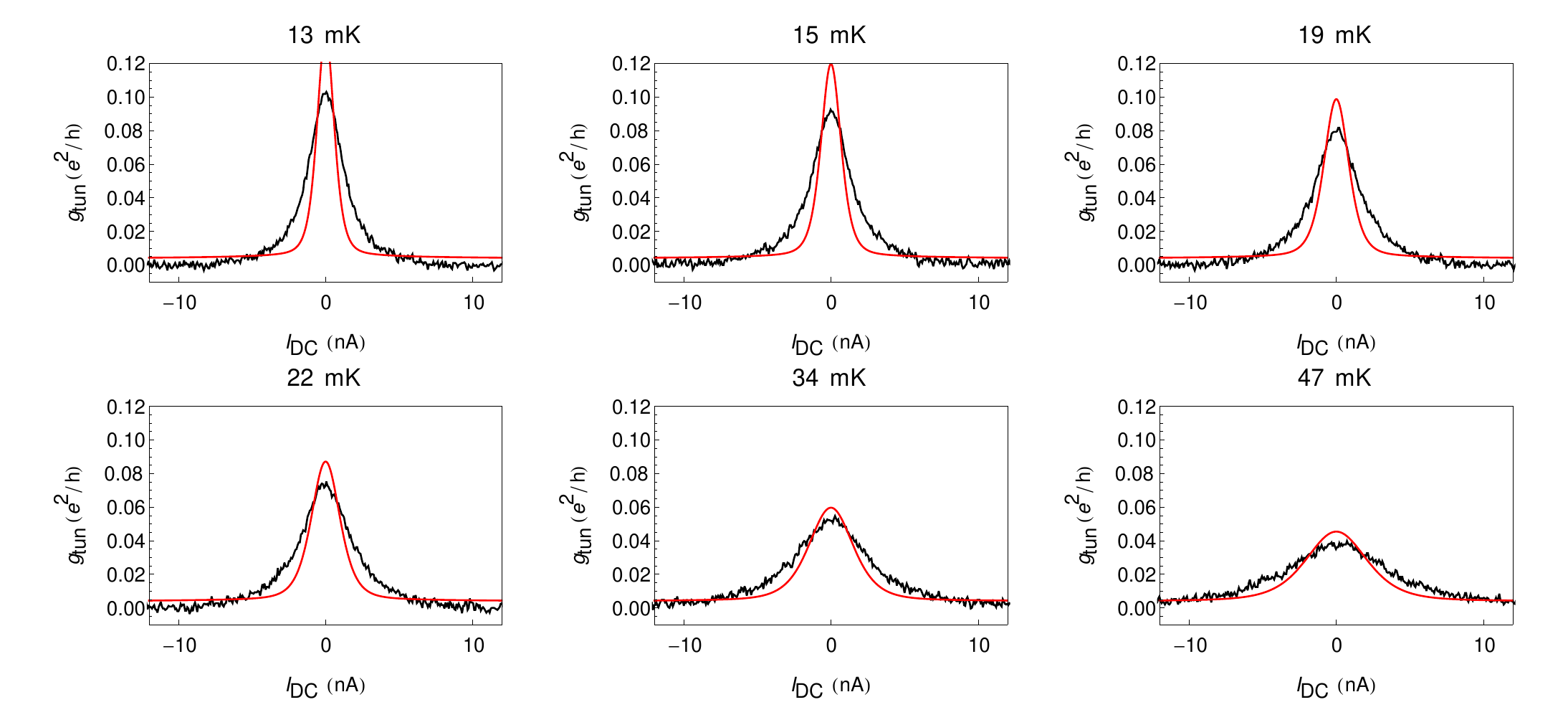}\\
	\vspace{-0.1cm}
	$\nu$ = 8/3, $g$ = 7/24, $e^*/e$ = 1/3\\
	\includegraphics[width=10cm]{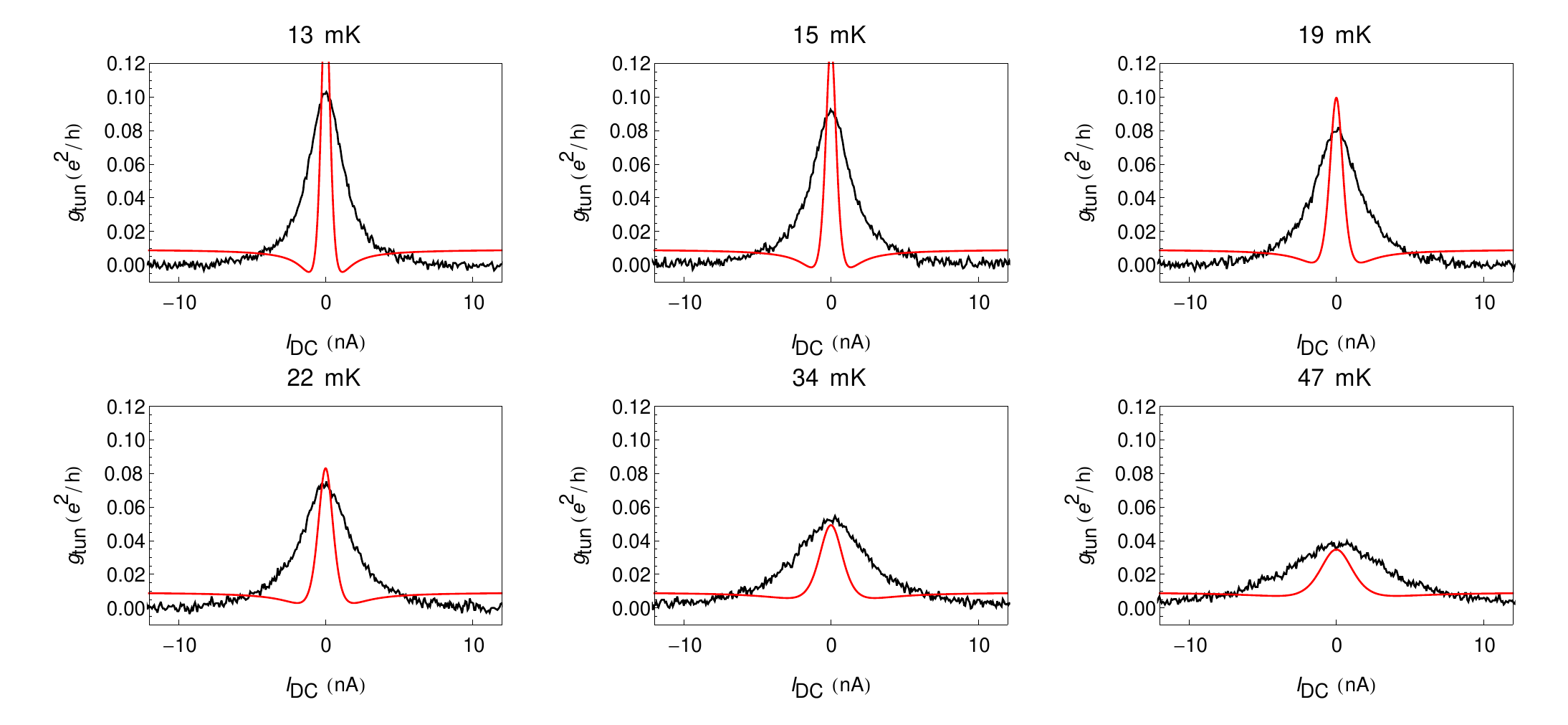}\\
		\vspace{-0.1cm}
	$\nu$ = 8/3, $g$ = 1/6, $e^*/e$ = 1/6\\
	\includegraphics[width=10cm]{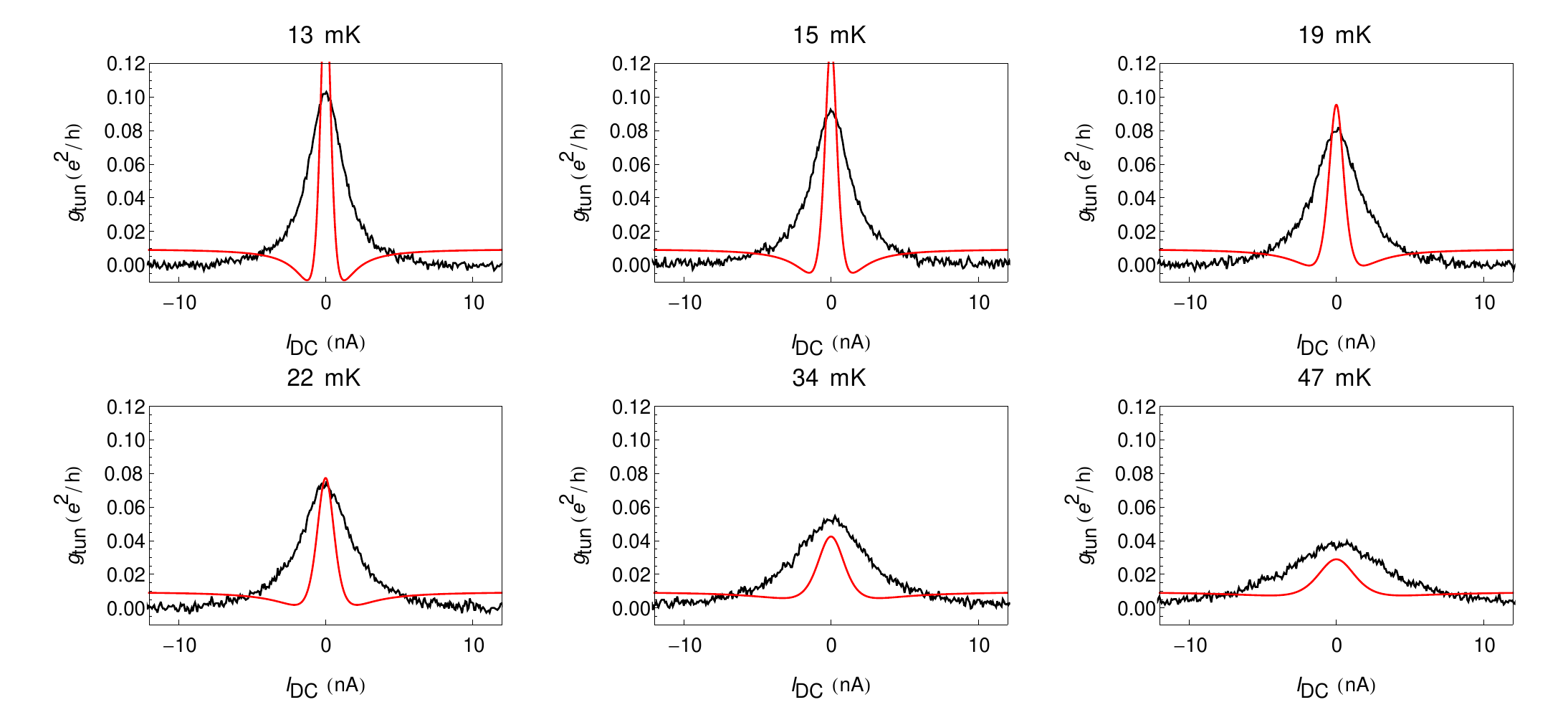}\\
	\end{center}
	\caption{\vspace{-1cm} Comparison between experimental and calculated $g_\mathrm{tun}$ for proposed parameter pairs  ($\nu$ = 8/3).}
	\label{Para83}
	\end{figure}

\end{document}